\documentclass{article}
\usepackage{arxiv}
\usepackage{graphicx,amsmath,amsfonts,amscd,amsthm,amssymb,pstricks}
\usepackage{color}% Include figure files
\usepackage{dcolumn}% Align table columns on decimal point
\usepackage{bm}% bold math
\usepackage{mathrsfs}
\usepackage{textcomp}
\usepackage{esvect}
\usepackage{float}
\usepackage[USenglish,british]{babel}
\usepackage{environ}
\usepackage{xifthen}
\usepackage{xargs}		% programming: better newcommand
\usepackage{hyperref}
\usepackage{physics,mathtools}
\usepackage[separate-uncertainty = true]{siunitx}
\usepackage{multirow}
\usepackage{comment}
\usepackage[caption=false]{subfig}

\newcommand{\ie}{i.e.\ }

\newcommand{\ham}{\mathcal{H}}
\newcommand{\eps}{\varepsilon}
\newcommand{\pr}{P}
\DeclareMathOperator{\diag}{diag}
\def\block(#1,#2)#3{\multicolumn{#2}{c}{\multirow{#1}{*}{$ #3 $}}}
\allowdisplaybreaks[1]

\newtheorem*{theorem*}{Theorem}
\newtheorem*{lemma*}{Lemma}

\title{Cleaning the beam halo using nonlinear AC magnets}

\author{F. Capoani\\
Dipartimento di Fisica e Astronomia, Universit\`a di Bologna and INFN Bologna, via Irnerio 46, Bologna, Italy\\
\And
A. Bazzani\\
Dipartimento di Fisica e Astronomia, Universit\`a di Bologna and INFN Bologna, via Irnerio 46, Bologna, Italy\\
\And
M. Giovannozzi\thanks{Corresponding author: massimo.giovannozzi@cern.ch}\\
Beams Department, CERN, Esplanade\ des Particules 1, 1211 Geneva 23, Switzerland}

\begin{document}
\maketitle

\begin{abstract}
Recently, nonlinear effects have been utilized to cool a special beam represented by an annular distribution in a 2D phase space. This outcome was accomplished using an AC dipole combined with amplitude detuning generated by static nonlinear magnets. In this paper, we investigate a more realistic scenario in which a beam distribution in a 4D phase space includes the presence of a beam halo and demonstrate how the latter can be removed using nonlinear effects. The proposed approach employs high-order nonlinear AC magnets to trap into nonlinear resonances and to perform an adiabatic transport of the beam halo in phase space. Theoretical models will be formulated and examined through numerical simulations to evaluate their efficacy.
\end{abstract}

%\pacs{29.20.db,29.27.Bd,05.45.-a,45.05.+x}
%\keywords{}444\maketitle must follow title, authors, abstract, \pacs, and \keywords
\section{Introduction}

The study of nonlinear effects highlights new phenomena in beam dynamics that could pave the way for innovative beam manipulation strategies. An example is adiabatic trapping into a stable nonlinear resonance to shape the transverse beam distribution~\cite{Capoani:2863096}. This concept underpins the beam splitting used in CERN Multi-Turn Extraction (MTE)~\cite{PhysRevLett.88.104801,PhysRevSTAB.7.024001,PhysRevSTAB.12.014001}, which has been a routine procedure in the operation of the CERN Proton Synchrotron for several years~\cite{Borburgh:2137954,PhysRevAccelBeams.20.014001,PhysRevAccelBeams.20.061001}.

Yet, this is not the only nonlinear manipulation possible. Inspired by~\cite{PhysRevLett.110.094801}, evidence suggests that it is possible to control the redistribution of invariants between the two transverse degrees of freedom~\cite{our_paper7}, on the condition that a suitable two-dimensional nonlinear resonance is traversed. This suggests new methods in manipulating the transverse beam emittances.

Recently, nonlinear effects have been shown to be highly successful in achieving cooling for a distinctive 2D beam~\cite{PhysRevAccelBeams.26.024001}, characterized by an annular distribution. This distribution is represented by a non-zero density within a range of radii $r_1 < r < r_2, \; r_1 > 0$ in the normalized phase space. This distribution can be produced by applying a single transverse kick to a centered beam under the condition of decoherence. The principal component for the proposed cooling technique is an AC dipole, which influences the dynamics of the particles by imparting a time-dependent deflection, irrespective of the actual coordinates in the phase space. For this reason, the beam distribution in the phase space was assumed to be devoid of particles to prevent any potential influence of the AC dipole on the action of the initial conditions near the origin of the phase space.

In this paper, we extend this methodology to devise a technique to remove the beam halo. The key concept here is that of high-order AC magnets. Conceptually, this generalizes the AC dipole, allowing for a phase-space region around the origin where the AC element has minimal or no impact on beam dynamics. Thus, the high-order nonlinear AC magnets can be used to perform an adiabatic trapping of particles in the beam halo while leaving the core particles unaffected. This approach could become a promising technique based on the removal of the halo particles. We remark that, at this stage of our research, the feasibility and performance of these AC magnets have not been considered.

Although it is evident that this method is not yet as advanced as current beam-halo control techniques in circular particle accelerators, such as the Hollow Electron Lens (HEL)~\cite{PhysRevLett.107.084802,stancari:napac13-tuoca1,redaelli:ipac15-webb1,PhysRevAccelBeams.23.031001,PhysRevAccelBeams.24.021001,Redaelli_2021}, it appears to be a notable and promising avenue for future developments.

The structure of the paper is as follows: A detailed investigation of a Hamiltonian system that includes an AC nonlinear magnet is presented in Section~\ref{sec:gen}, with a focus on the 2D scenario (one degree of freedom) and its potential application to beam-halo cleaning. The 2D Hamiltonian model is used to make some analytical computations and propose a protocol to manipulate the beam halo. The map models, derived from the prototype Hamiltonian, used for the numerical analysis of this beam-halo cleaning method are discussed in Section~\ref{sec:model}. The map models are more realistic than the Hamiltonian ones and allow probing the wealth of phenomena that might occur in an accelerator ring. Extensive numerical simulations of these models are conducted to assess the technique, and the results are elaborated in Section~\ref{sec:numres}. Finally, the conclusions are drawn in Section~\ref{sec:conc}. The detailed mathematics behind the derivation of the Hamiltonian model and the discussion of its fixed points are included in the Appendices~\ref{sec:app} and~\ref{sec:app1}, respectively.
\section{A Hamiltonian model for the proposed beam-halo manipulation} \label{sec:gen}
Horizontal betatron motion in the presence of an AC multipole can be described by the Hamiltonian of a generic oscillator with sextupole nonlinearity and $2q$-polar time-dependent excitation~\cite{PhysRevSTAB.5.054001, peggstang, bei1999beam}, namely,
\begin{equation}
\begin{split}
\ham(x,p_x,t) & = \ham_0(x, p_x) + \ham_n(x,t) \\
\ham_0(x, p_x) & = \omega_x \frac{x^2+p_x^2}{2} + \beta^{3/2}_x \frac{k_3}{3} x^3 \\
\ham_q(x,t) & = \eps_q \frac{x^q}{q} \cos\omega t \,  ,
\end{split}
\label{eq:ham_xp}
\end{equation}
where $x$ and $p_x$ are Courant-Snyder coordinates~\cite{Courant:593259}, $\beta_x$ is the horizontal beta function, $k_3$ is the sextupolar gradient, where the nonlinear gradient is generically defined as 
\begin{equation}
k_q = \frac{1}{B_0 \rho} \, \frac{\partial^{q-1} B_y}{\partial x^{q-1}} \, L \, ,
\end{equation}
where $B_0 \rho$ stands for the magnetic rigidity of the reference particle, $B_y$ is the transverse component of the magnetic field and $L$ is the physical length of the magnetic element. The parameter
\begin{equation}
    \eps_q = \beta^{q/2}_x k_q\, , 
\end{equation}
represents the strength of the nonlinear AC multipole.

The use of sextupole nonlinearity is somewhat arbitrary yet convenient because of the common presence in circular accelerator lattices. Octupoles could also have been utilized without altering the fundamental conclusions, as the role of nonlinearity is simply that of generating an amplitude-detuning term.

The case of an AC dipole ($q=1$) has been treated in~\cite{PhysRevAccelBeams.26.024001}. If $n>1$, we can still use the Normal Form approach to determine the nonresonant interpolating Hamiltonian of~\eqref{eq:ham_xp}~\cite{Bazzani:262179}) and expressing it in the action-angle coordinates $(\phi,J)$ of the unperturbed ($\eps_q=0$) system. In that case, the Hamiltonian reads, in the most general way, as
\begin{equation}
\begin{split}
\ham(\phi,J,t) & = \omega_x \, J + \sum_{l\ge 2} \frac{\Omega_l}{l} J^l + \\ 
               & + \eps_q \sum_{m,k} c_{q,m,k} J^{m/2} \cos (k\phi) \cos\omega t \, ,
\end{split}
\label{eq:hamgen}
\end{equation}
where we applied the Poincaré-Von Zeipel perturbation theory~\cite{turchetti} to Eq.~\eqref{eq:ham_xp} to find angle-action coordinates $(\phi, J$) and the detuning coefficients $\Omega_l$ and to write the expression of $x^q(\phi,J)$ as a Fourier expansion (details are found in the Appendix~\ref{sec:app}).

If we impose a $1:1$ resonant condition, \ie $\omega\approx\omega_x$, we can rewrite Eq.~\eqref{eq:hamgen} in a rotating resonance frame where the angle $\gamma=\phi-\omega t$ is a slow variable. The Hamiltonian reads 
\begin{equation}
\begin{split}
\ham(\gamma,J,\psi) & = (\omega_x -\omega) \, J + \sum_{l\ge 2} \frac{\Omega_l}{l} J^l \\
                    & + \eps_q\sum_{m,k}  c_{q,m,k} J^{m/2} \cos [k (\gamma + \psi)] \cos\psi \, ,
\end{split}
\label{eq:hamgen2}
\end{equation}
where $\psi=\omega t$. The average over the fast variable $\psi$ is non-zero only for $k=1$. Therefore, we can define $c_{q,m}=c_{q,m,1}$, remove the sum over $k$, and truncate our perturbative expansion to the order $J^{\hat{m}/2}$ for which the AC term gives its first non-zero contribution. In that case, Eq.~\eqref{eq:hamgen} reduces to
\begin{equation}
\begin{split}
\ham(\gamma,J) & = (\omega_x - \omega) \, J + \sum_{l=2}^{\hat{m}/2} \frac{\Omega_l}{l} J^l + \eps_q c_{q,\hat{m}} J^{\hat{m}/2} \cos\gamma \, .
\end{split}
\end{equation}

For a sextupolar nonlinearity, the first detuning coefficient is $\Omega_2=-5\, \beta^3_x \, k_3^2/(6 \, \omega_x)$. We then compute $c_{q,m}$ as 
\begin{equation}
c_{q,m} = \frac{1}{2\pi q} \Re \int_0^{2\pi} \dd\phi\, e^{i\phi} [x^q]_{m/2} \, ,
\end{equation}
where $[z]_u$ stands for the coefficient of the $J^u$ term in the expression of $z$.

Details on the calculation of $\hat m$ and $c_{q,\hat m}$ for different values of $q$ are given in Appendix~\ref{sec:app}. Here, we summarize the results for some interesting cases:
\begin{equation}
\begin{split}
q=3: \quad \ham (\gamma,J) & =(\omega-\omega_x)J + \frac{\Omega_2}{2}J^2 + \\
                        & + \eps_3 \, c_{3,3} J^{3/2}\cos\gamma \,, \\
q=4: \quad \ham (\gamma,J) & =(\omega-\omega_x)J + \frac{\Omega_2}{2}J^2 + \\
                        & + \eps_4 \, c_{4,5} J^{5/2}\cos\gamma \,, \\
q=5: \quad \ham (\gamma,J) & =(\omega-\omega_x)J + \frac{\Omega_2}{2}J^2 + \\
                        & + \eps_5 \, c_{5,5} J^{5/2}\cos\gamma \,, \\
q=6: \quad \ham (\gamma,J) & =(\omega-\omega_x)J + \frac{\Omega_2}{2}J^2 + \frac{\Omega_3}{3}J^3 + \\ 
                        & + \eps_6 \, c_{6,7} J^{7/2}\cos\gamma \\
q=7: \quad \ham (\gamma,J) & =(\omega-\omega_x)J + \frac{\Omega_2}{2}J^2 + \frac{\Omega_3}{3}J^3 + \\ 
                        & + \eps_7 \, c_{7,7} J^{7/2}\cos\gamma \,. 
\end{split}
\label{eq:ham-models}
\end{equation}
An octupole or a decapole ($q=4$ or $q=5$) AC element are represented by the same power of $J$ (as well as dodecapole and the $14$-pole), and we note that the Hamiltonian description for $q=6, 7$ includes a higher-order term of amplitude detuning.

Furthermore, we note that the case of the sextupole ($q=3$) is less interesting for applications because it would affect the chromaticity and the low-field region around the origin would be less extended than for higher-order AC multipoles. Hence, we will focus on the cases with $q=4$ or $q=5$, whose difference mainly lies in the different expression of the value of the Fourier coefficient that multiplies the AC multipole amplitude $\eps_q$ (see Eq.~\eqref{eq:Fourier}). It should be stressed that the expressions of the two coefficients are completely different, since $c_{4,5}$ depends on $\omega_x, k_3$ and $\beta_x$, while $c_{5,5}$ is a numerical constant. This characteristic can be used as a selection criterion for a suitable choice of the AC multipole.

\begin{figure}[htp]
    \centering  \includegraphics[trim=7truemm 7truemm 14truemm 7truemm,width=0.5\columnwidth,clip=]{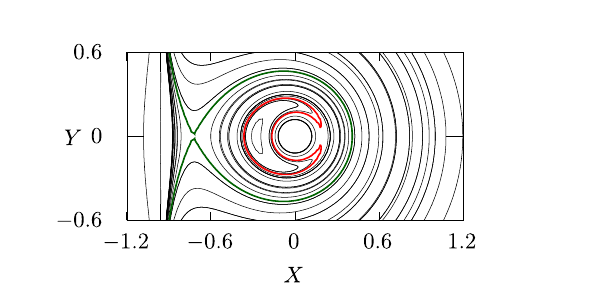}\\
    \includegraphics[trim=7truemm 7truemm 14truemm 7truemm,width=0.5\columnwidth,clip=]{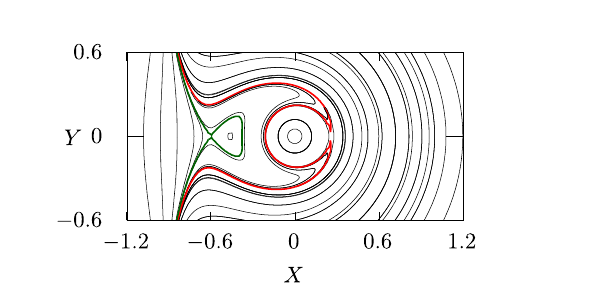}\\
    \includegraphics[trim=7truemm 1truemm 14truemm 7truemm,width=0.5\columnwidth,clip=]{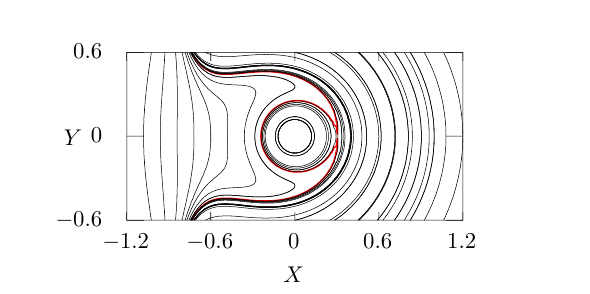}
    \caption{Phase-space portraits of the Hamiltonian~\eqref{eq:hamxy} with parameters $\lambda=-0.1$ (top), $\lambda=-0.18$ (center), $\lambda=-0.25$ (bottom), $\mu_5=1$. The red and the dark green lines represent the separatrices.}
    \label{fig:ham_phsp0}
\end{figure}

The next steps consist of performing an analysis of the phase-space topology of the Hamiltonian models~\eqref{eq:ham-models}. It is convenient to introduce two parameters $\lambda, \mu_q$
\begin{equation}
\lambda = 4 \frac{\omega-\omega_x}{\Omega_2} \qquad \text{and} \qquad \mu_q = 2^\frac{3-m}{2} \frac{\eps_q \, c_{q,m}}{\Omega_2} \, , 
\end{equation}
which encode the distance from resonance and the multipole strength, respectively. The Hamiltonians of Eq.~\eqref{eq:ham-models} ($q=4$ or $q=5$) can be rescaled and cast in the following form 
\begin{equation}
\begin{split}
    \ham (X,Y,q) & = \lambda (X^2+Y^2) + (X^2+Y^2)^2 + \\
                 & + \mu_q (X^2+Y^2)^2 X \\
                 & = (X^2+Y^2) \left [ \lambda + (X^2+Y^2) \left ( 1 + \mu_q X \right ) \right ] \, , 
\end{split}
\label{eq:hamxy}
\end{equation}
where $X=\sqrt{2J}\cos\gamma$, $Y=\sqrt{2J}\sin\gamma$. We observe that $\ham$ is invariant under the transformation $(X, Y) \to (X, -Y)$.  Furthermore, the origin $(X=0,Y=0)$ is always a stable fixed point because the multipole has no effect at zero amplitude. It is easy to show that there are no fixed points along the $Y$-axis. On the other hand, solving $\pdv*{\ham}{X}=0$ for $Y=0$ we can find up to three other fixed points on the $X$-axis. The resulting cubic equation
\begin{equation}
    5 \mu_q X^3 + 4 X^2 + 2 \lambda = 0 \, , 
    \label{eq:cubic}
\end{equation}
in fact, has three real solutions if $-128/(675\mu_q^2) < \lambda < 0$. In this case, the phase-space portrait has two possible configurations, which are shown in the top and center plots of Fig.~\ref{fig:ham_phsp0}, while the bottom represents a case with $\lambda < -128/(675\mu_q^2)$ and a single real solution of Eq.~\eqref{eq:cubic}.

We also note that no other fixed points are present and this result is discussed in the Appendix~\ref{sec:app1}.

It is also possible to compute an explicit closed-form expression $Y=Y(X)$ for the level lines of the Hamiltonian~\eqref{eq:hamxy} and observe that singularities might be present, in the sense that $Y(X) \to \infty$ when $X \to -1/\mu_q$. This behavior is clearly visible in the phase-space portraits in Fig.~\ref{fig:ham_phsp0}, where there is a singularity at $X=-1$, as expected. 

\begin{figure}[htp]
    \centering    \includegraphics[trim=7truemm 6truemm 5truemm 8truemm,width=0.5\columnwidth,clip=]{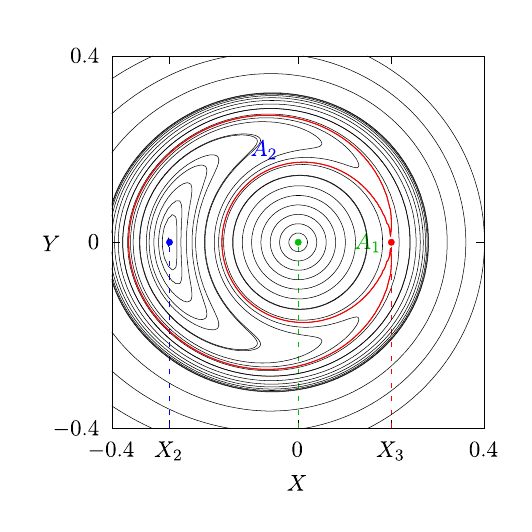}
    \caption{Phase-space portrait of the Hamiltonian~\eqref{eq:hamxy} with parameters $\lambda=-0.1$, $\mu_5=1$ zoomed in to the area closer to the origin. The separatrix is represented in red and the fixed points and the phase-space regions are identified.}
    \label{fig:ham_phsp}
\end{figure}

We are particularly interested in the regime corresponding to $\lambda \lesssim 0$ and the phase-space structures at small amplitude because, in the map model that we employ in the numerical simulations (see the next section), the high-amplitude fixed points and separatrices are outside of the dynamic aperture of the map. We therefore focus on the phase-space structure of Fig.~\ref{fig:ham_phsp}, which is similar to the one considered in Ref.~\cite{PhysRevAccelBeams.26.024001}, with a central region ($A_1$) whose stable fixed point is at the origin and a c-shaped island ($A_2$) surrounding it, with a stable fixed point at $(X_2, 0)$ and a hyperbolic point at $(X_3, 0)$, whose positions can be evaluated numerically.

\begin{figure}[htp]
    \centering   \includegraphics[trim=0truemm 1truemm 0truemm 1truemm,width=0.6\columnwidth,clip=]{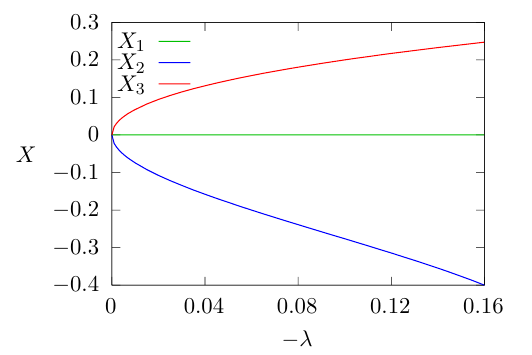}
    \includegraphics[trim=-3truemm 0truemm 0truemm 1truemm,width=0.6\columnwidth,clip=]{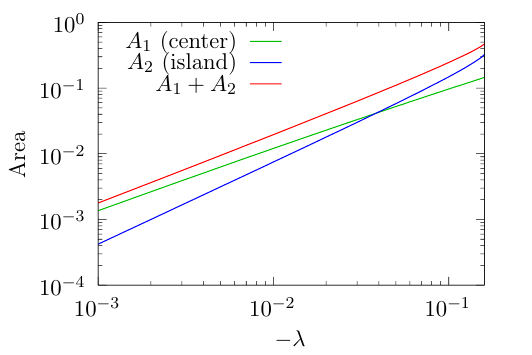}
    \caption{Position of fixed points (top) and area of the central region ($A_1$) and of the island ($A_2$) (bottom) of the Hamiltonian~\eqref{eq:hamxy} for $\mu_5=1$, as a function of $\lambda$ in the interval where the phase-space topology is equivalent to that of Fig.~\ref{fig:ham_phsp}. The areas under consideration depend on $\lambda$ almost like a power law.}
    \label{fig:fp_area}
\end{figure}

It is also possible to evaluate the area of $A_1$ and $A_2$ for a fixed value of $\mu_q$ as a function of $\lambda$, which is shown in Fig.~\ref{fig:fp_area} (top). The bifurcation of the fixed points is clearly visible and $\lambda$ can be used to control their distance to the origin, thus controlling the area in phase space where the motion is quasi-linear. Concerning the area of $A_1$ and $A_2$, it is clearly visible in the bottom panel of the same figure that they are constantly increasing as a function of $\lambda$ with a variation very close to a power law.

\section{Beam-halo cleaning using nonlinear AC magnets} \label{sec:model}

\subsection{The principle of the proposed method}

The method of removing the beam halo is based on the generation of an adiabatic time variation of the system parameters $\lambda, \mu_q$, which in turn induces a variation of the position of the fixed points and of the area of the regions $A_1$ and $A_2$. This enables adiabatic trapping and transport of initial conditions in phase space, and upon application of an appropriate time variation of the parameters, similarly to the cooling of an annular beam distribution~\cite{PhysRevAccelBeams.26.024001}, it is possible to trap high-amplitude particles and move them towards even higher amplitude in a controlled way. By doing so, the particles will either be lost in an aperture limit of the physical system, e.g., a collimator jaw, or lost because of the dynamic aperture. In either case, the high-amplitude particles, which are part of the beam halo, will be removed from the beam distribution. 

Figure~\ref{fig:fp_area} (bottom) shows that both $A_1$ and $A_2$ increase their area as $\lambda$ moves farther from the resonance. This means that if $\lambda$ is adiabatically changed, according to Neishtadt's separatrix crossing theory~\cite{neish1975}, A particle that orbits in phase space with action $J$ outside the regions $A_1$ and $A_2$, will be trapped in $A_1$ or $A_2$ in a probabilistic way, when $\lambda=\lambda^*$ and $A_1(\lambda^*)+A_2(\lambda^*)=2\pi J$. Indeed, the trapping into resonance occurs in a random way with probabilities that depend on the time derivative of the areas of $A_1$ and $A_2$ according to
\begin{equation}
\pr_i =\begin{dcases}
\quad 1 &\quad\text{ if } \xi_i > 1\\
\quad\xi_i &\quad\text{ if } 0<\xi_i < 1\\
\quad 0 &\quad\text{ if } \xi_i < 0
\end{dcases}\, ,
\end{equation}
where
\begin{equation}
    \xi_i = {\frac{\dv*{A_i}{\lambda}}{\dv*{(A_1+A_2)}{\lambda}}}\eval_{\lambda=\lambda^*} \qquad i=1,\,2 \, ,
\end{equation}
which is always different from zero. 

The trapped particle assumes an action $J=2\pi A_i(\lambda^*)$, which means that a generic particle orbiting outside of $A_1$ and $A_2$ will be trapped in the resonance or in the central region: in the first case, it will then be transported towards a higher amplitude to be removed eventually from the beam distribution; in the latter case, its action will be reduced so that it will no longer be in the halo region but rather in the core part of the beam distribution.

The goal of our approach is to remove from the beam distribution the particles with amplitude higher than a certain value, which corresponds to the qualitative concept of a beam halo, using the effect of the AC multipole. Therefore, we switch on the multipole at a frequency that creates the island structure at the needed amplitude, and move $\lambda$ away from the resonance to trap a number of high-amplitude particles into the island region and then transport them at an even higher amplitude (or even out of the dynamical aperture of the system), where they can be removed from the beam distribution without harming the core of the beam.

Some particles will be trapped in the center region, at a lower but still high value of the amplitude, as $A_1$ also increases during the process. However, since the process is essentially probabilistic, one can simply iterate the procedure many times to ensure that a large enough fraction of the beam halo is removed. The procedure should be carried out leaving the beam emittance of the core beam unaffected during the trapping and transport of the high-amplitude particles.

\subsection{2D map model}

To numerically test the proposed approach of beam-halo removal, we first build a model that, although close to the Hamiltonian models of Eq.~\eqref{eq:ham-models}, better reproduces the transverse dynamics in a circular accelerator. To this aim, we consider a 2D Hénon-like map with sextupolar and octupolar nonlinearities that generate the amplitude detuning, and an oscillating multipole magnet ($q=4$ for octupole or $q=5$ for a decapole), which gives the following polynomial map
\begin{equation}
\begin{pmatrix} x_{n+1} \\ p_{x,n+1}  \end{pmatrix} =  \bm{R}(\omega_x)  \begin{pmatrix} x_n \\ p_{x,n} + \Delta p_{x,n,q}  \end{pmatrix}
\label{eq:map2d}
\end{equation}
where $\bm{R}(\omega)$ is a $2\times 2$ rotation matrix and the nonlinear kick $\Delta p_{x,n,q}$ is given by
\begin{equation}
\begin{split}
    \Delta p_{x,n,q} & = \beta_x^{3/2} {k_3} x_n^2 + \beta_x^2 {k_4} x_n^3 + \\
                   & + \beta_x^{q/2} { \eps_q} x_n ^q\cos n \omega =\\
                   &= \hat k_3 x_n^2 + \hat k_4 x_n^3 + \hat \eps_q x_n ^q\cos n \omega \,,
\end{split}
\end{equation}
where we introduced the scaled parameters $\hat k_3$, $\hat k_4$, $\hat \eps_q$.

\begin{figure}
\centering
\includegraphics[trim=7truemm 7truemm 3truemm 7truemm,width=0.6\columnwidth,clip=]{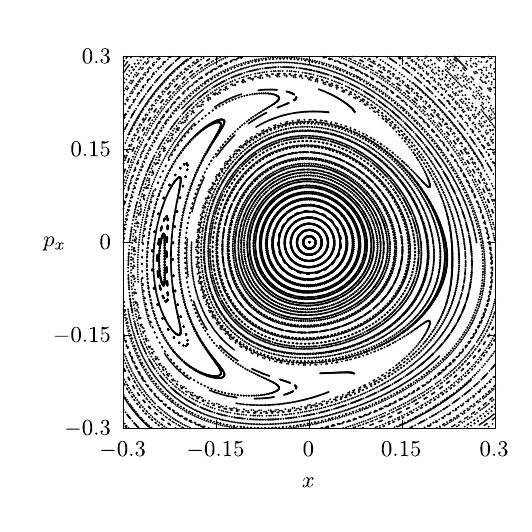}
\caption{Phase-space portrait of the map of Eq.~\eqref{eq:map2d}, with $\omega_x=0.414$, $\omega=0.413$, $k_3=2$, $k_4=0$, $\eps_5=0.2$.}
\label{fig:map_phsp}
\end{figure}

For $\omega \approx \omega_x$, we expect that the phase space generated by the map is topologically equivalent to that generated by the Hamiltonian~\eqref{eq:hamxy} and shown in Fig.~\ref{fig:ham_phsp}. This is indeed the case, as can be seen in Fig.~\ref{fig:map_phsp}, even though some chains of stable higher-order fixed points are visible, which are not present in the phase-space portrait of the Hamiltonian. This is a consequence of the non-integrable nature of the map (\ref{eq:map2d}) with many nonlinear resonances in the phase space.

\subsection{4D map model}

The polynomial map of Eq.~\eqref{eq:map2d} can be extended to simulate the full 4D transverse space including nonlinear coupling effects between the $x$ and $y$ planes. We consider a 4D Hénon-like map with nonlinearities and an oscillating normal multipole magnet (octupole or decapole, as before)
\begin{equation}
\begin{pmatrix} x_{n+1} \\ p_{x,n+1} \\ y_{n+1} \\ p_{y,n+1} \end{pmatrix} = \begin{pmatrix} R(\omega_{x,n}) & 0 \\ 0 & R(\omega_{y,n}) \end{pmatrix} \begin{pmatrix} x_n \\ p_{x,n} + \Delta p_{x,n,q} \\ y_n \\ p_{y,n} + \Delta p_{y,n,q} \end{pmatrix}
\end{equation}
where $R(\omega)$ are $2\times 2$ rotation matrices and the nonlinear kicks $\Delta p_{x,n,q}$ and $\Delta p_{y,n,q}$ are given by
\begin{equation}
\begin{split}
    \Delta p_{x,n,q} &= \beta_x^{3/2}{k_3} (x_n^2- \chi y_n^2) \\&+ \beta_x^2 {k_4}(x_n^3-3\chi x_ny_n^2) \\&+ \beta_x^{q/2} { \eps_q} \Re(x_n + i\chi^{1/2} y_n)^q\cos n \omega 
\end{split}
\end{equation}
and
\begin{equation}
\begin{split} 
    \Delta p_{y,n,q} = & - 2 \beta_x^{3/2} \chi k_3 x_n y_n \\& - \beta_x^2 \chi {k_4}(\chi y_n^3-3 x_n^2y_n) \\&- \beta_x^{q/2} \chi^{1/2} {\eps_q} \Im(x_n + i\chi^{1/2}y_n)^q\cos n \omega \, , 
\end{split}
\end{equation}
where $\chi = \beta_y/\beta_x$. Also in this case, it is possible to introduce scaled parameters $\hat k_3$, $\hat k_4$, $\hat \eps_q$ with the same definition as in the 2D case, \ie the scaling includes only powers of the beta functions, but not $\chi$. 

It is worth noting that the 4D model used in the numerical simulations is more complicated than a simple generalization of the 2D map. First, the nonlinear kicks can be represented in a generic way by
\begin{equation} 
\mathbf{K}\,:\, \begin{pmatrix} x \\ p_x \\ y \\ p_y \end{pmatrix} \mapsto \begin{pmatrix} x \\ p_x + \sqrt{\beta_x} f_{x}(\sqrt{\beta_x} x, \sqrt{\beta_y} y) \\ y \\ p_y + \sqrt{\beta_y} f_{y}(\sqrt{\beta_x} x, \sqrt{\beta_y} y) \end{pmatrix} \, , 
\end{equation}
when in the original physical coordinates $(x, p_x, y, p_y)$ are given by 
\begin{equation} 
\mathbf{\hat K}\,:\, \begin{pmatrix} x \\ p_x \\ y \\ p_y \end{pmatrix} \mapsto \begin{pmatrix} x \\ p_x + f_{x}(x, y) \\ y \\ p_y + f_{y}(x, y) \end{pmatrix} \, .
\end{equation}

We finally consider a model of an accelerator lattice of length $L$, whose Twiss parameters at $s=0$ are given by $\beta_x = \beta_{x,0}$, $\beta_y=\beta_{y,0}$, $\alpha_x = \alpha_y = 0$ and at $s=L/2$ they are given by $\beta_x = \beta_{y,0}$, $\beta_y=\beta_{x,0}$, $\alpha_x=\alpha_y=0$. For the sake of symmetry, two sources of multipolar nonlinearities are located at $s=0$ and $s=L/2$. Furthermore, to optimize the overall performance of the 4D system, two nonlinear AC elements are placed at $s=0$ and $s=L/2$. Note that $\chi (0)=1/\chi(L/2)$, for this reason, at the location where $\chi > 1 $ an exciter of normal type is installed, while at the location where $\chi < 1$ the exciter is of skew type. This allows the exciters to operate under almost 2D conditions, the normal one affecting mainly the horizontal phase space, and the skew one the vertical phase space.

The one-turn map of this lattice can be written as
\begin{equation}
\mathbf{X}' = \mathbf{M}_2 \mathbf{\hat{K}}_2 (\, \mathbf{M}_1 \mathbf{\hat{K}}_1 (\mathbf{X})) \, ,    
\end{equation}
where $\mathbf{M}_{1,2}$ represent the transfer matrices of the lattice sections in between the nonlinearities. If we introduce the normalization matrix 
\begin{equation}
\mathbf{T}(s) = \diag({\beta_x}^{1/2}(s), {\beta_x}^{-1/2}(s), {\beta_y}^{1/2}(s), {\beta_y}^{-1/2}(s)) 
\end{equation}
(assuming $\alpha_x=\alpha_y=0$), then, we can write the one-turn map of the full accelerator lattice in normalized coordinates according to
\begin{equation}
\begin{split}
    \mathbf{X}' & =  \mathbf{M}_2 \mathbf{\hat{K}}_2 \left (\, \mathbf{T}(\tfrac{L}{2}) \mathbf{T}^{-1}(\tfrac{L}{2}) \mathbf{M}_1 \mathbf{T}(0) \mathbf{T}^{-1}(0) \mathbf{\hat{K}}_1 (\mathbf{T}(0) \mathbf{x}) \right ) \\ 
                & = \mathbf{M}_2 \mathbf{\hat{K}}_2 \left (\, \mathbf{T}(\tfrac{L}{2})\mathbf{R}(\tfrac{\bm\omega}{2}) \mathbf{K}_1 (\mathbf{x}) \right ) \, ,     
\end{split}
\label{eq:map0}
\end{equation}
which has been obtained by means of the following relationships
\begin{align}
\mathbf{R}(\tfrac{\bm\omega}{2}) & = \mathbf{T}^{-1}(\tfrac{L}{2}) \mathbf{M}_1 \mathbf{T}(0) \label{eq:transf1} \\
\mathbf{K}_1 (\mathbf{x}) & = \mathbf{T}^{-1}(0) \mathbf{\hat{K}}_1 (\mathbf{T}(0) \mathbf{x}) \label{eq:transf2} \, , 
\end{align}
where $\mathbf{R}$ is a rotation matrix and the nonlinear kick has the following expression 
\begin{equation} 
\mathbf{K}_1 \,:\, \begin{pmatrix} x \\ p_x \\ y \\ p_y \end{pmatrix} \mapsto \begin{pmatrix} x \\ p_x + \sqrt{\beta_{x,0}} f_{x,1}(\sqrt{\beta_{x,0}} x, \sqrt{\beta_{y,0}} y) \\ y \\ p_y + \sqrt{\beta_{y,0}} f_{y,1}(\sqrt{\beta_{x,0}} x, \sqrt{\beta_{y,0}} y) \end{pmatrix} \, . 
\end{equation}

The map~\eqref{eq:map0} can then be recast in the following form
\begin{equation}
\begin{split}
    \mathbf{x}' & = \mathbf{T}^{-1}(0) \mathbf{M}_2 \mathbf{T}(\tfrac{L}{2}) \mathbf{T}^{-1}(\tfrac{L}{2}) \mathbf{\hat{K}}_2 (\, \mathbf{T}(\tfrac{L}{2}) \mathbf{R}(\tfrac{\bm\omega}{2}) \mathbf{K}_1 (\mathbf{x})) \\
    & = \mathbf{R}(\tfrac{\bm\omega}{2}) \mathbf{K}_2 (\, \mathbf{R}(\tfrac{\bm\omega}{2}) \mathbf{K}_1 (\mathbf{x})) \, ,     
\end{split}
\label{eq:map1}
\end{equation}
where we have made use of the following relationships
\begin{align}
\mathbf{R}(\tfrac{\bm\omega}{2}) & = \mathbf{T}^{-1}(0) \mathbf{M}_2 \mathbf{T}(\tfrac{L}{2}) \label{eq:transf3} \\
\mathbf{K}_2 (\mathbf{x}_1) & = \mathbf{T}^{-1}(\tfrac{L}{2}) \mathbf{\hat{K}}_2 \left (\mathbf{T}(\tfrac{L}{2}) \mathbf{x}_1 \right ) \label{eq:transf4}
\end{align}
and, using $\mathbf{x}_1 = \mathbf{R}(\tfrac{\bm\omega}{2})\mathbf{K}_1(\mathbf{x})$, we have 
\begin{equation} 
\mathbf{K}_2 \,:\, \begin{pmatrix} x_1 \\ p_{x1} \\ y_1 \\ p_{y1} \end{pmatrix} \mapsto \begin{pmatrix} x_1 \\ p_{x1} + \sqrt{\beta_{y,0}} f_{x,2}(\sqrt{\beta_{y,0}} x_1, \sqrt{\beta_{x,0}} y_1) \\ y_1 \\ p_{y1} + \sqrt{\beta_{x,0}} f_{y,2}(\sqrt{\beta_{y,0}} x_1, \sqrt{\beta_{x,0}} y_1) \end{pmatrix} \, 
\end{equation}
In the complex notation $z=x+iy$, the general expression of the four kick functions reads:
\begin{equation}
\begin{split}
f_{x,1}(z) &= \phantom{-}\Re \left ( k_3 z^2 + k_4 z^3 + \eps_q z^q \cos n\omega \right ) ,\\
f_{y,1}(z) &=  -\Im \left ( k_3 z^2 + k_4 z^3 + \eps_q z^q \cos n\omega \right ) ,\\
f_{x,2}(z) &= \phantom{-}\Re \left ( k_3 z^2 + k_4 z^3 + i\eps_q z^q \cos n\omega \right ) ,\\
f_{y,2}(z) &=  -\Im \left ( k_3 z^2 + k_4 z^3 + i\eps_q z^q \cos n\omega \right ) .\\
\end{split}
\label{eq:kicks1}
\end{equation}

Note that the presence of the imaginary unit that multiplies the term $\eps_q z^q \cos n\omega$ makes the second AC element a skew element.

The numerical simulations have been performed using the general from expressed in Eq.~\eqref{eq:kicks1}, but also the variant form given by
\begin{equation}
\begin{split}
f_{x,1}(z) &= \phantom{-}\Re \left ( k_3 z^2 + k_4 z^3 + \eps_q z^q \cos n\omega \right ) ,\\
f_{y,1}(z) &=  -\Im \left ( k_3 z^2 + k_4 z^3 + \eps_q z^q \cos n\omega \right ) ,\\
f_{x,2}(z) &= \phantom{-}\Re \left ( ik_3 z^2 + ik_4 z^3 + i\eps_q z^q \cos n\omega \right ) ,\\
f_{y,2}(z) &=  -\Im \left ( ik_3 z^2 + ik_4 z^3 + i\eps_q z^q \cos n\omega \right ) .\\
\end{split}
\label{eq:kicks2}
\end{equation}

The reason for this choice is that the kicks in the form of Eq.~\eqref{eq:kicks1} generate an asymmetric amplitude detuning, much stronger in $x$ than in $y$, due to the factor $\chi(0) > 1$. The kicks in the form of Eq.~\eqref{eq:kicks2}, instead, generate a symmetric amplitude detuning, due to the presence of the skew sextupole and octupole. 

\begin{figure}
\centering
\includegraphics[trim=0truemm 5truemm 0truemm 2truemm, width=0.7\textwidth,clip=]{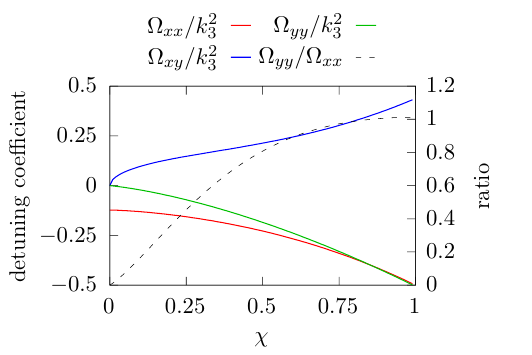} \\
\includegraphics[trim=0truemm 0truemm 0truemm 12.5truemm, width=0.7\textwidth,clip=]{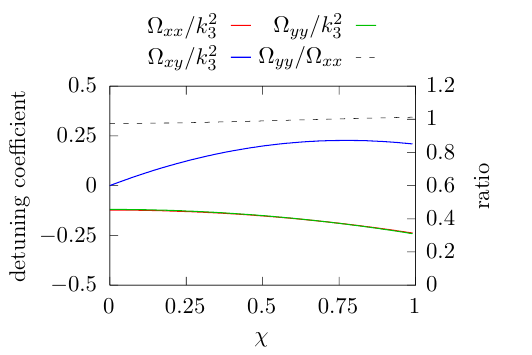}
\caption{Coefficients of the amplitude detuning computed using the Normal Form Hamiltonian $H=\omega_x J_x + \omega_y J_y + \Omega_{xx} J_x^2/2 + \Omega_{xy} J_x J_y + \Omega_{yy} J_y^2 /2$ obtained by applying the algorithm of Ref.~\cite{Bazzani:262179,bazzani1995199} to the map described in Eq.~\eqref{eq:kicks1} (top) and Eq.~\eqref{eq:kicks2} (bottom), with $\omega_x=0.414$ and $\omega_y=0.424$, as function of $\chi =\beta_{y,0}/\beta_{x,0}$. Note that for the map of Eq.~\eqref{eq:kicks2} $\Omega_{xx}=\Omega_{yy}$ and the amplitude-detuning coefficients are perfectly symmetric.}
\label{fig:detuning}
\end{figure}

We recall that the tune $\Omega_x, \Omega_y$ can be expressed as
\begin{equation}
    \begin{split}
    \Omega_x(J_x,J_y) & =\omega_x + \Omega_{xx} J_x + \Omega_{xy} J_y \\
    \Omega_y(J_x,J_y) & =\omega_y + \Omega_{xy} J_x + \Omega_{yy} J_y \, , 
    \end{split}
\end{equation}
and that $\Omega_{xx}, \Omega_{xy}, \Omega_{yy} \propto k_3^2$ for our model~\cite{bazzani1995199}. The characteristics of the amplitude-detuning coefficients are evident in Fig.~\ref{fig:detuning}, which shows these coefficients calculated using Normal Forms~\cite{Bazzani:262179,bazzani1995199} for the two maps. The symmetry between $\Omega_{xx}$ and $\Omega_{yy}$ in the map of Eq.~\eqref{eq:kicks2} is clearly shown in the bottom graph. It is important to note that in our simplified model, the inclusion of a skew element is crucial to achieve symmetric amplitude detuning behavior. However, in an actual accelerator lattice, this symmetry can be accomplished with standard magnets placed in suitable sections of the regular arcs (an example being the LHC lattice~\cite{LHCDR}, where the Landau octupoles, used to mitigate collective instabilities, create a symmetric amplitude detuning in the two transverse planes).

\section{Results of numerical simulations} \label{sec:numres}

\begin{figure}[htb]
\centering
\includegraphics[trim=0truemm 2truemm 0truemm 5truemm,width=0.5\columnwidth,clip=]{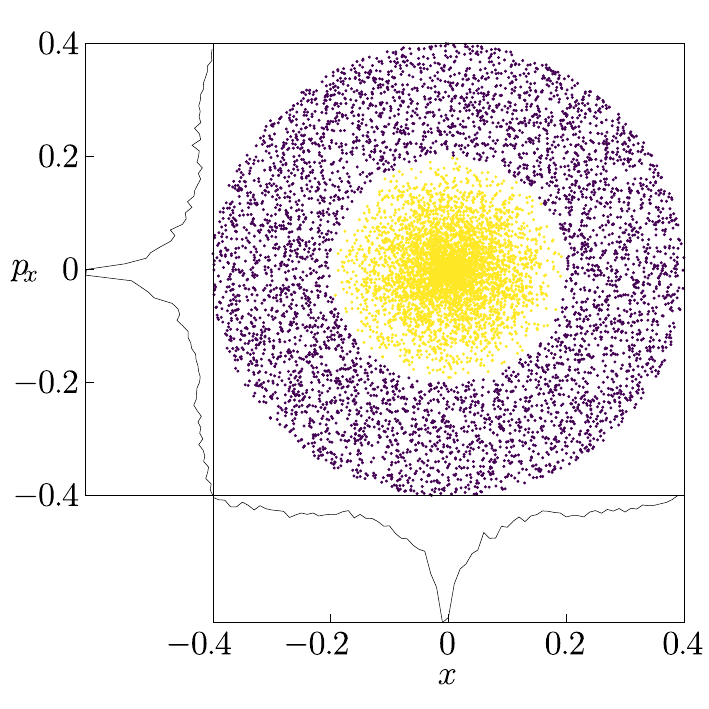}
\includegraphics[trim=0truemm 2truemm 0truemm 5truemm,width=0.5\columnwidth,clip=]{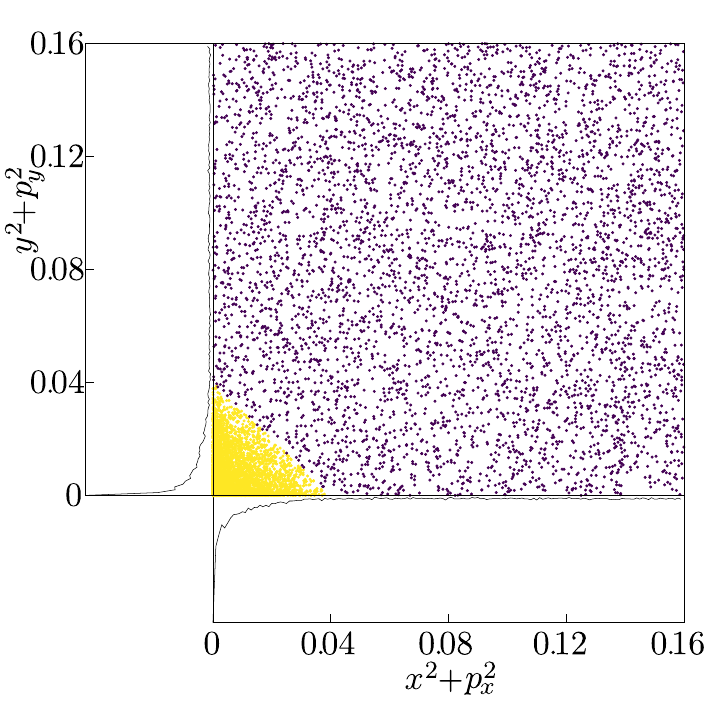}
\caption{Initial conditions in the $(x,p_x)$ phase space and in the action space for the 2D (top) and 4D (bottom) cases used in the numerical simulations. The yellow conditions represent the beam core, the violet ones the beam halo. On each plot, the smaller plots represent, in arbitrary units, the projection of the distribution on each axis.}
\label{fig:incond}
\end{figure}

\subsection{Distribution of initial conditions}

The initial conditions $(x,\,y,\,p_x,\,p_y)$ in the 4D phase space are generated according to a given distribution which depends only on the amplitude variable $r=\sqrt{x^2 + p_x^2 +  y^2 + p_y^2}$, the angular part of the distribution being uniform. Depending on their amplitude, we classify particles as in the core or in the halo of the beam, defining two quantities, $r_1$ and $r_2$ with $0<r_1<r_2$. The particles with $0< r < r_1$ identify the \textit{ beam core}, whereas the particles with $r_1 < r < r_2$ identify the \textit{beam halo}. We also assume that particles with an amplitude greater than $r_2$ are removed by the beam by the action of collimators or other aperture limits.

To perform the numerical simulations, we generate $n_1$ initial conditions in the core and $n_2$ in the halo, with $n_1=n_2=\num{5e3}$. The distribution in the core follows a $q$-Gaussian function~\cite{curado1992} with $q=0$ and maximum radius $r_1$, while the halo is modeled as a uniform annular distribution in the $r_1<r<r_2$ region. The two distributions are:
\begin{equation}
    \begin{split}
    \rho_\text{c}(r; r_1) & = \mathcal{N}_\text{c} e_0\qty(-\frac{r^2}{r_1^2}), \qquad r \in [0, r_1] \\
    \rho_\text{h}(r; r_1, r_2) & = \mathcal{N}_\text{h} \frac{r}{r_2^2 - r_1^2},\qquad r\in ]r_1,r_2]
    \end{split}
\end{equation}
where $\mathcal{N}_\text{h,c}$ are normalization constants and $e_q(r)$ is the $q$-exponential function given by
\begin{equation}
e_q(r) = \left [ 1 + (1 - q) r\right]^\frac{1}{1-q} \, .
\end{equation}

The choice of a $q$-Gaussian with $q=0$ for the distribution of the core is based on the fact that it does not have tails and is zero for $r=r_1$, which allows easily defining the edge of the core. Of course, for the case of 2D numerical simulations, we use the same approach by simply setting $y=p_y=0$. 

An example of the distributions used is shown in Fig.~\ref{fig:incond}, where the 2D distribution (top) is shown in the $(x, p_x)$ phase space, while the 4D distribution (bottom) is represented in the linear action space. The yellow and violet markers represent the core and halo distributions, respectively. The projections of the distribution along the axis are also shown. The clear separation between the core and the halo, made possible by the choice of the $q$-Gaussian distribution, is clearly visible.

\subsection{2D map model}

\begin{figure}
\centering
\includegraphics[trim=1truemm 0truemm 5truemm 2truemm, width=0.6\textwidth,clip=]{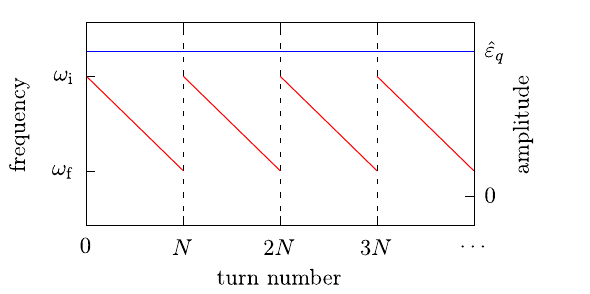}
\includegraphics[trim=4truemm 0truemm -13truemm 2truemm, width=0.6\textwidth,clip=]{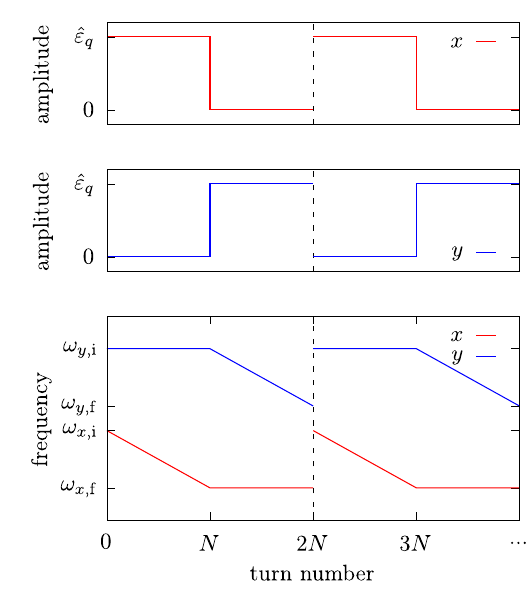}
\caption{Description of the modulation protocol of the AC multipole used for the 2D (top) and 4D (bottom) in the numerical simulations, showing the strength of the AC multipoles and their frequencies as a function of the number of turns. The vertical dashed lines highlight the times when the particles with radius higher than $r_2$ are removed from the beam distribution. In the 4D case, the cleaning is performed in each plane sequentially, to minimize the coupling between horizontal and vertical motion.}
\label{fig:process}
\end{figure}

The assessment of the performance of the proposed approach to clean the beam halo has been carried out by studying the 2D case first. The numerical simulations are performed using the 2D map model of Eq.~\eqref{eq:map2d}, changing the amplitude and frequency of the AC multipole according to the protocol illustrated in Fig.~\ref{fig:process} (top). The main goal is to determine, as a function of the value of the main parameters of the system, namely $\omega_x, \omega_\mathrm{i}, \eps_q$ and $N \gg 1$, the elapsed time during the parameter change, which fraction of particles is removed from the beam halo. Additionally, it is important to assess whether the beam core is affected by the manipulations, either in terms of intensity loss or emittance growth. Therefore, in the plots with the simulation results, we will represent the conditions that preserve at least 99\% of the intensity of the beam core with a solid line, while a dotted line is used when a larger fraction of the beam core is lost.

We study 2D map models with two multipoles, namely an AC octupole ($q=4$) and an AC decapole ($q=5$). The scans of the main parameters are performed using some `default values for parameters that are not part of the scan study, and these values are: $\hat\eps_4=\hat\eps_5=0.2$, $N=\num{1e5}$, $\omega_x=0.414$, $\omega_\text{i}=0.413$, $\omega_\text{f}=0.407$, $\hat k_3=1$, $\hat k_4=0$. In addition, the beam-halo cleaning process is repeated $10$ times, according to the scheme shown in Fig.~\ref{fig:process} (top), and the performance is evaluated at the end.

\begin{figure*}
\centering
\subfloat[\label{fig:2dsim_oct-a}]{%
\includegraphics[trim=1truemm 1truemm 1truemm 2truemm, width=0.47\columnwidth,clip=]{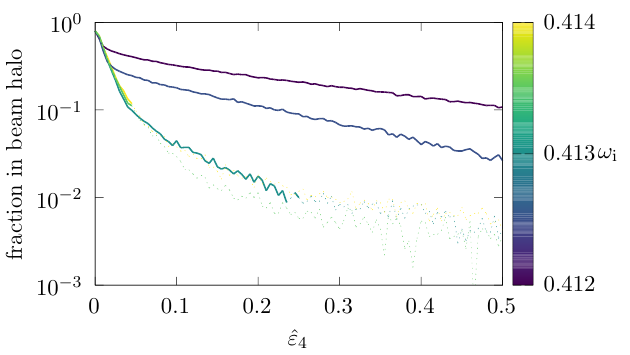}}
\subfloat[\label{fig:2dsim_dec-a}]{%
\includegraphics[trim=1truemm 1truemm 1truemm 2truemm, width=0.47\columnwidth,clip=]{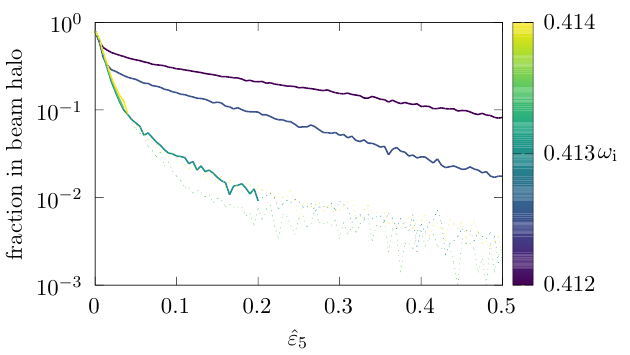}} \\
\subfloat[\label{fig:2dsim_oct-b}]{%
\includegraphics[trim=1truemm 1truemm 1truemm 2truemm, width=0.47\columnwidth,clip=]{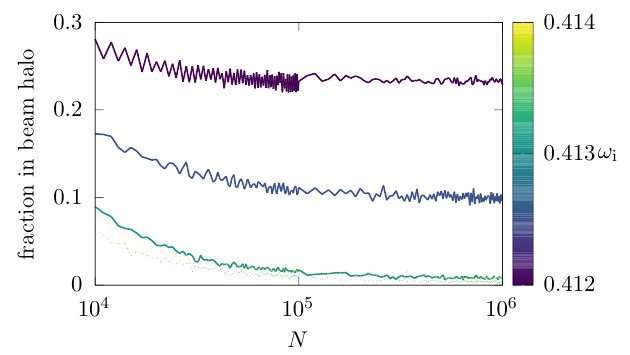}}
\subfloat[\label{fig:2dsim_dec-b}]{%
\includegraphics[trim=1truemm 1truemm 1truemm 2truemm, width=0.47\columnwidth,clip=]{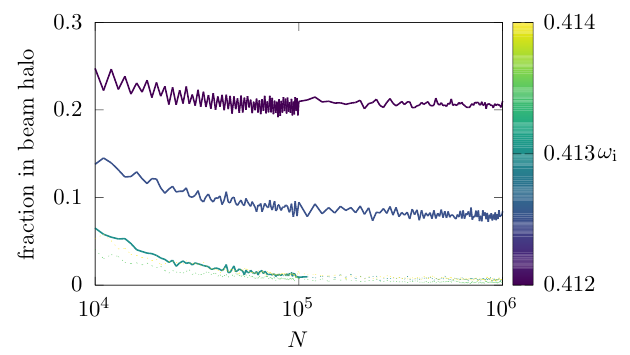}} \\
\subfloat[\label{fig:2dsim_oct-c}]{%
\includegraphics[trim=1truemm 1truemm 1truemm 2truemm, width=0.47\columnwidth,clip=]{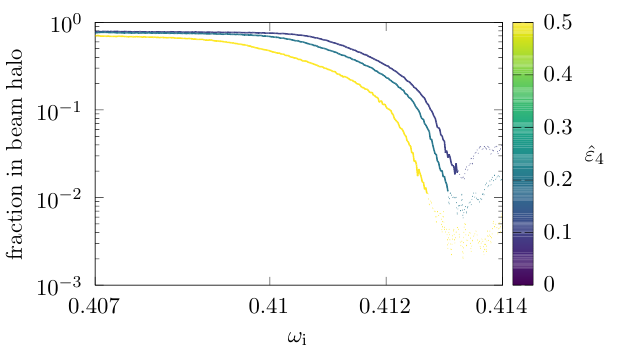}} 
\subfloat[\label{fig:2dsim_dec-c}]{%
\includegraphics[trim=1truemm 1truemm 1truemm 2truemm, width=0.47\columnwidth,clip=]{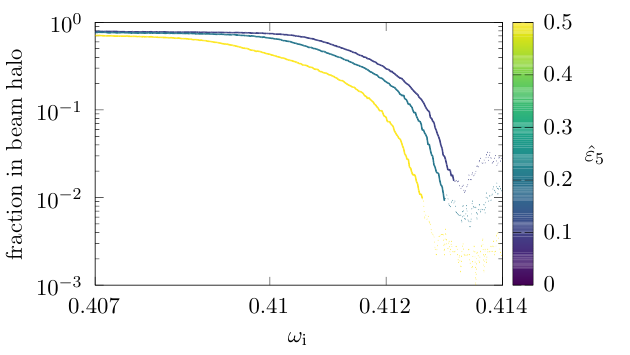}} \\
\subfloat[\label{fig:2dsim_oct-d}]{%
\includegraphics[trim=1truemm 1truemm 1truemm 2truemm, width=0.47\columnwidth,clip=]{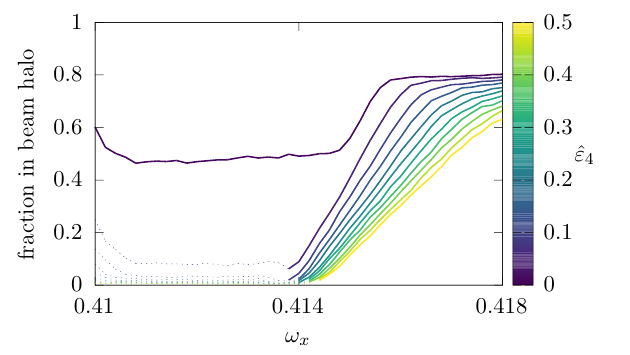}} 
\subfloat[\label{fig:2dsim_dec-d}]{%
\includegraphics[trim=1truemm 1truemm 1truemm 2truemm, width=0.47\columnwidth,clip=]{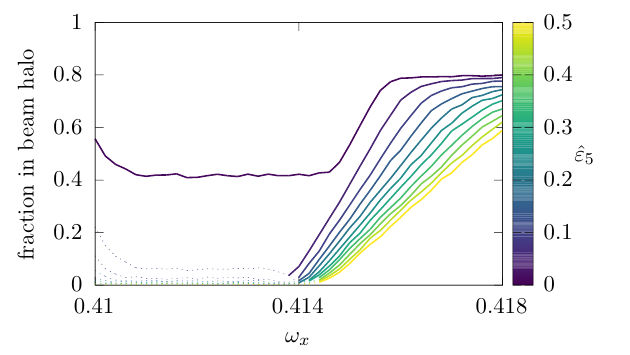}}
\caption{Survival fraction of particles of the beam halo after the cleaning procedure using the 2D model of the map in Eq.~\eqref{eq:map2d} with $q=4$ (left column) or with $q=5$ (right column). The dependence on the AC multipole strength (first row), number of turns (second row), initial AC multipole frequency (third row), and beam tune (fourth row) is probed, following the protocol of Fig.~\ref{fig:process} (top). The dotted lines represent conditions for which more than $1\%$ of the beam core is lost during the process. Unless otherwise stated, we used $\hat \eps_4=0.2$ (or $\hat \eps_5=0.2$), $N=\num{1e5}$, $\omega_x=0.414$, $\omega_\text{i}=0.413$, $\omega_\text{f}=0.407$, $\hat k_3$ = 1, $\hat k_4 = 0$ and the beam-halo cleaning process is repeated ten times.}
\label{fig:2dsim}
\end{figure*}

In Fig.~\ref{fig:2dsim} (left column), we show the results for the octupole case, while those for the decapole case are shown in the right column. Globally, the results are very similar for both two AC multipoles. Figures~\ref{fig:2dsim_oct-a} and~\ref{fig:2dsim_dec-a} represent the fraction that remains in the beam halo as a function of the AC multipole strength $\hat\eps_q$. Our findings indicate that in both scenarios, increasing the strength of the AC multipole enhances beam-halo removal. However, this also raises the potential for losses within the beam core, particularly as the initial AC multipole frequency near the $1:1$ resonance. However, in both cases, it is feasible to achieve a removal of the 99\% halo while limiting the loss of the beam-core intensity to 1\%.

Figures~\ref{fig:2dsim_oct-b} and~\ref{fig:2dsim_dec-b} illustrate how the fraction of particles remaining in the halo depends on the number $N$ of turns used to adjust the AC multipole frequency, thus assessing the impact of adiabaticity of the process on the overall efficiency of beam-halo cleaning. Naturally, higher values of $N$ lead to an increased trapping efficiency. However, this also raises the likelihood of trapping particles from the core within the island, thereby removing them from the beam. The graphs indicate that the default value of $N=\num{1e5}$ is nearly optimal, as the beam-halo removal rate exhibits a minimal change between $N=\num{1e5}$ and $N=\num{1e6}$.

Figures~\ref{fig:2dsim_oct-c} and~\ref{fig:2dsim_dec-c} illustrate how halo cleaning depends on the initial AC multipole frequency $\omega_\text{i}$ for several values of $\hat\eps_q$. These results serve as a complement to those shown in Figs.~\ref{fig:2dsim_oct-a} and~\ref{fig:2dsim_dec-a}. It is notable that even when $\omega_\mathrm{i}$ is significantly distant from the resonance value and from the region where the resonance island is formed, far from the center and the most crucial part of the halo, a small portion of the beam halo can still be cleaned using a very high AC multipole strength. However, in these cases, the particles are more likely to be removed by hitting the dynamic aperture rather than through a controlled trapping and transport mechanism. As $\omega_\mathrm{i}$ gets closer to $\omega_x$, which is the resonance condition, the island forms closer to the center. An optimal value of the frequency can be found that allows a larger fraction of the halo to be trapped in successive iterations of the protocol, with a minimal impact on the beam core.

Lastly, in Figs.~\ref{fig:2dsim_oct-d} and~\ref{fig:2dsim_dec-d}, we illustrate the effects when the primary frequency $\omega_x$ deviates from the standard value of $\omega_x=0.414$. In both octupole and decapole cases, a slight increase in $\omega_x$ (setting $\omega_\text{i}=0.413$ and thus moving further from resonance) decreases the efficiency of the process. In contrast, reducing $\omega_x$, which brings it closer to resonance, results in unacceptable losses in the core.

\begin{figure}
\centering
\includegraphics[trim=0truemm 0truemm 0truemm 2truemm, width=0.7\columnwidth,clip=]{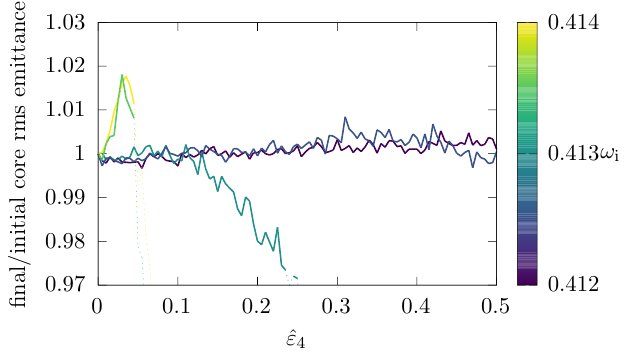}
\includegraphics[trim=0truemm 0truemm 0truemm 2truemm, width=0.7\columnwidth,clip=]{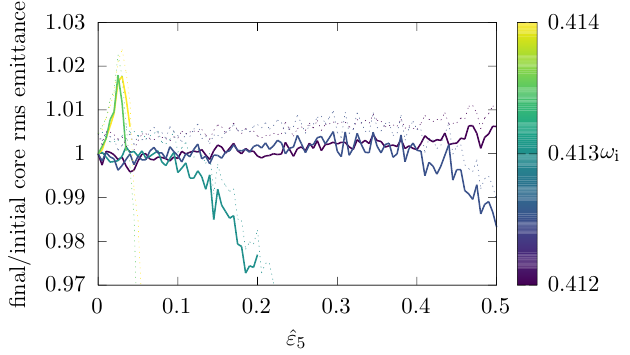}
\caption{Ratio between final and initial rms emittance for the octupole (top) and decapole (bottom) model as a function of $\hat\eps_q$ in the 2D map model of Eq.~\eqref{eq:map2d}. The initial emittance is defined as the beam rms emittance after $\num{1e6}$ turns performed with $\hat\eps_q=0$. This allows computing the impact of filamentation due to the nonlinearities used to the generate the amplitude detuning. Note that the filamentation effects generate an rms emittance increase of $0.46\%$ compared to the rms emittance of the ensemble of initial conditions. We used $\omega_x=0.414$, $\omega_{\text{f}}=0.407$, $N=\num{1e5}$, $\hat k_3=1$, $\hat k_4=0$.}
\label{fig:2dsim_emitt}
\end{figure}

As previously stated, eliminating the beam halo and preserving the emittance of the beam core are the two crucial requirements for the efficiency of the process. In Fig.~\ref{fig:2dsim_emitt} we show the ratio between the rms emittance $\epsilon = \sqrt{\det \mathrm{Cov}(x,p_x)}$ of the core particle ensemble at the end of the cleaning process and before the cleaning process starts, as a function of $\hat\eps_4$ (top) and $\hat\eps_5$ (bottom) for different values of $\omega_\text{i}$. To consider potential filamentation phenomena resulting from the nonlinearities introduced in the model to generate the amplitude detuning, initial numerical simulations track the particle distribution for $N=\num{1e6}$ turns with $\eps_q=0$. At this stage, the core emittance is assessed and used to normalize the final core emittance post-cleaning. The normalized emittance has increased by a factor of $1.0046$ compared to the initial core emittance. As shown in Fig.~\ref{fig:2dsim}, the conditions where the beam loss exceeds $1\%$ are marked with a dotted line. For both AC multipoles, specific values of $\omega_\mathrm{i}$ can be found where the emittance remains unaffected by the cleaning process within a certain strength range of the AC multipoles. In contrast, significant particle losses result in increased core emittance because of the transport of particles to the outer regions by means of resonance islands. However, in other scenarios, a significant decrease in core emittance is noted as particles are transported in a controlled manner to the halo region and subsequently removed by the cleaning mechanism.

\subsection{4D map model: asymmetric amplitude detuning}
The studies on halo manipulation have been extended to a comprehensive 4D transverse model of particle dynamics using the map outlined in Eq.~\eqref{eq:map1} and the setup described in Eq.~\eqref{eq:kicks1}. Two types of AC multipoles ($q=4$ or $q=5$) are alternately activated following the protocol depicted in Fig.~\ref{fig:process} (bottom) executed $10$ times. The strategy of alternating the activation of two multipoles with reversed $\chi$ values is intended to trap and transport the beam halo to higher amplitudes, ensuring independent removal in each plane. However, the nonlinear coupling introduced by an AC multipole in the orthogonal plane can impact manipulation and degrade overall performance. Thus, it is crucial to examine the influence of the coefficient $\chi$, which allows each AC multipole, both normal and skew, to function in a quasi-2D manner.

Due to the features of the variation in amplitude detuning across the two planes, it is necessary to apply different amplitudes for each exciter. Thus, a crucial parameter in the model is the ratio between the amplitudes of the normal multipole and the skew multipole. The standard parameters for the 4D map~\eqref{eq:map1} are: $\hat k_3=1$, $\beta_x=1$, and $\hat k_4=0$, with the following parameters remaining constant unless otherwise specified: $\omega_x=0.414$, $\omega_y=0.424$, $\omega_{x,\text{i}}=0.413$, $\omega_{y,\text{i}}=0.423$, $\omega_{x,\text{i}}=0.407$, $\omega_{y,\text{i}}=0.417$, $\hat\eps_4=\hat\eps_5=0.1$, $N=\num{1e5}$.

\begin{figure*}
\centering
\subfloat[\label{fig:4dsim-oct-a}]{%
\includegraphics[trim=0truemm 3truemm 3truemm 13truemm, width=0.38\columnwidth,clip=]{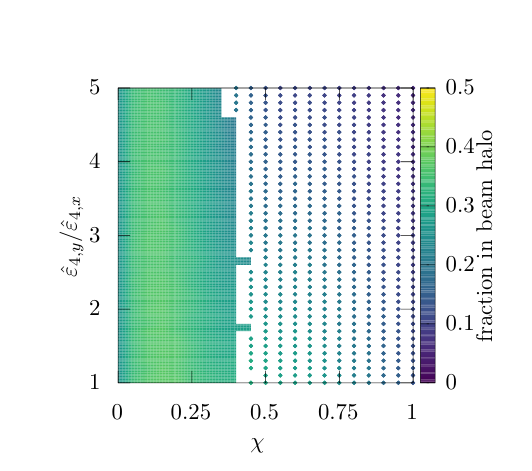}} %\hspace{20truemm}
\subfloat[\label{fig:4dsim-dec-a}]{%
\includegraphics[trim=0truemm 3truemm 3truemm 13truemm, width=0.38\columnwidth,clip=]{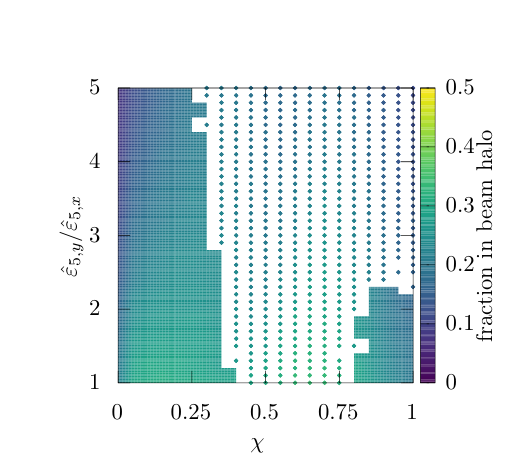}}\\\hspace{20truemm}
\subfloat[\label{fig:4dsim-oct-b}]{%
\includegraphics[trim=1truemm 1truemm 3truemm 2.5truemm, width=0.38\columnwidth,clip=]{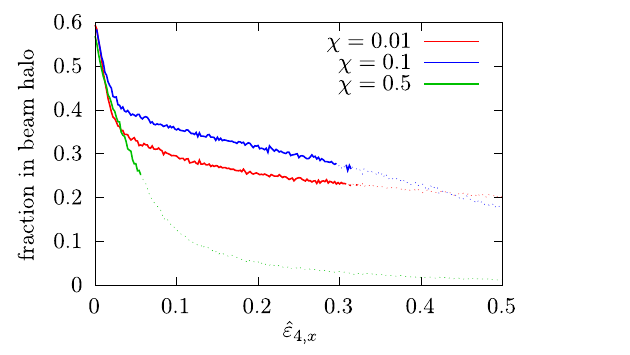}} %\hspace{20truemm} % nuova figura inserita
\subfloat[\label{fig:4dsim-dec-b}]{%
\includegraphics[trim=1truemm 1truemm 3truemm 2.5truemm, width=0.38\columnwidth,clip=]{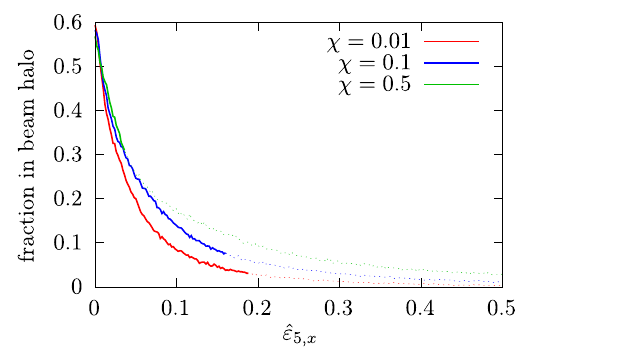}} \\ % nuova figura inserita
\subfloat[\label{fig:4dsim-oct-c}]{%
\includegraphics[trim=0truemm 3truemm 3truemm 13truemm, width=0.38\columnwidth,clip=]{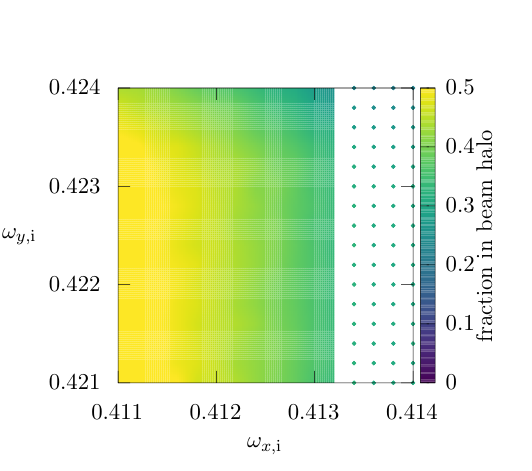}} %\hspace{20truemm}
\subfloat[\label{fig:4dsim-dec-c}]{%
\includegraphics[trim=0truemm 3truemm 3truemm 13truemm, width=0.38\columnwidth,clip=]{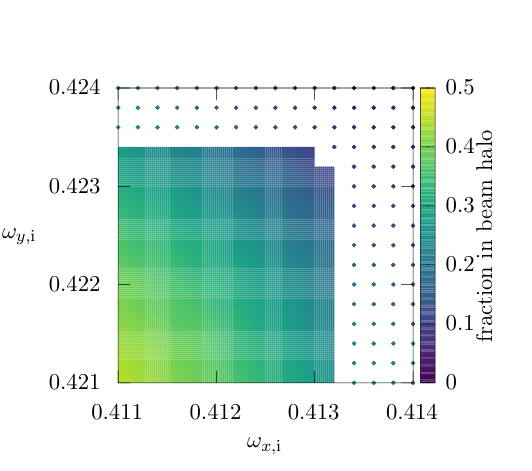}} \\
\subfloat[\label{fig:4dsim-oct-d}]{%
\includegraphics[trim=1truemm 3truemm 3truemm 13truemm, width=0.38\columnwidth,clip=]{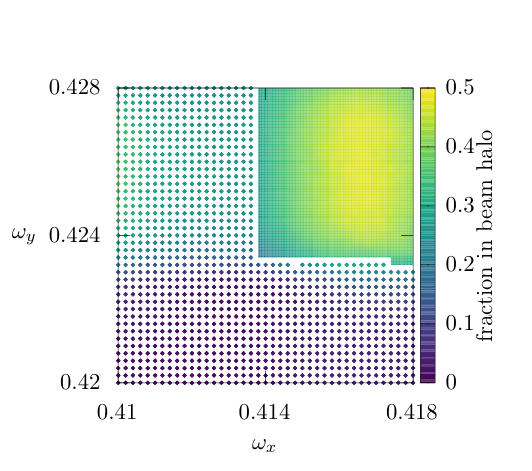}} %\hspace{20truemm}
\subfloat[\label{fig:4dsim-dec-d}]{%
\includegraphics[trim=1truemm 3truemm 3truemm 13truemm, width=0.38\columnwidth,clip=]{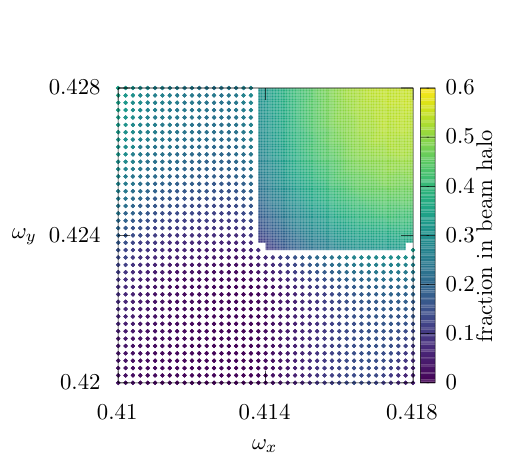}}
\caption{Survival fraction of particles of the beam halo after the cleaning procedure using the 4D model of the map in Eq.~\eqref{eq:map1}, and the configuration of Eq.~\eqref{eq:kicks1}, with $q=4$ (left column) or with $q=5$ (right column). Unless otherwise stated, we used $\hat k_3=1$, $\hat k_4=0$, $\hat\varepsilon_{q} =0.1$, $\hat\varepsilon_{q,y} = 5\hat\varepsilon_{q,x}$, $\chi=0.1$, $\omega_x=0.414$, $\omega_y=0.424$, $\omega_{x,\text{i}}=0.413$, $\omega_{x,\text{f}}=0.407$, $\omega_{x,\text{f}}=0.423$, $\omega_{y,\text{f}}=0.417$, $N=\num{1e5}$ and the beam-halo cleaning process is repeated $10$ times. In the 2D histograms, the filled regions represent conditions that preserve more than $99\%$ of the beam-core intensity.}
\label{fig:4dsim}
\end{figure*}

The results of numerical simulations to evaluate the efficiency of the proposed halo-cleaning method in 4D are shown in Fig.~\ref{fig:4dsim}. The left and right columns refer to the octupole and decapole scenarios, respectively.

Figures~\ref{fig:4dsim-oct-a} and~\ref{fig:4dsim-dec-a} present the efficiency of halo removal in the octupole and decapole models, correlated with the ratio of $\beta$ functions and the ration of the strength of the skew and normal AC multipoles. Generally, a value of $\chi<1/4$ ensures that the core intensity remains largely intact (we recall that in the plots, the color-gradient regions indicate configurations where the core intensity does not drop below 99\% of the initial value). Although the octupole model shows limited efficiency in halo reduction, the decapole model demonstrates superior performance, where a powerful skew decapole can eliminate over 90\% of the beam halo. Notably, the decapole model achieves remarkable results for $0.75 \leq \chi \leq 1$, effectively reducing the need for a strong skew AC decapole. However, we will not consider this particular configuration in any further detail. Instead, we will proceed with the assumption of a 5:1 ratio between the skew and normal AC elements, as previously indicated.

Figures~\ref{fig:4dsim-oct-b} and~\ref{fig:4dsim-dec-b} present an adjusted version of previous studies, where we maintain the ratio $\hat\eps_{q,y}=5\hat\eps_{q,x}$ and show the proportion of particles remaining in the halo at the end of the process as a function of $\hat\eps_{q,x}$ for three values of $\chi$. Strong AC multipoles significantly reduce the halo, but also excessively eliminate core particles, exceeding the acceptable limit of 1\%. The plots reveal the impact of selecting a value of $\chi$ outside the optimal small range, with substantial core losses occurring even at low $\hat\eps_q$. Additionally, the decapole case proves to be more efficient in halo removal. Particularly striking is the change in the slope of the curves concerning the AC octupole, suggesting a shift in the underlying dynamics, although a definitive explanation remains elusive.

In Figs.~\ref{fig:4dsim-oct-c} and~\ref{fig:4dsim-dec-c}, the effect of the AC multipole frequency on halo-removal efficiency is illustrated. The data show an increase in efficiency as the initial multipole frequency nears the resonance, a trend observed consistently in both transverse planes. However, the highest halo-removal efficiency also leads to a reduction in core intensity. Furthermore, for the octupole case, the dependence on $\omega_{x,\mathrm{i}}$ is more pronounced compared to $\omega_{y,\mathrm{i}}$. Conversely, in the decapole case, the frequency dependency is similar for both planes and relatively weak. Notably, the decapole outperforms the octupole, achieving superior halo-removal.
 
The influence on halo cleaning on the tunes of the 4D model is evident in Figs.~\ref{fig:4dsim-oct-d} and \ref{fig:4dsim-dec-d}. We observe that the core intensity remains well preserved, consistent with the 2D scenario, only when $\omega_x$ and $\omega_y$ diverge significantly from $\omega_{x,\text{i}}$ and $\omega_{y,\text{i}}$. Additionally, the overall behavior is quite similar for the two AC multipoles analyzed, with the decapole configuration consistently showing superior performance.

\begin{figure*}
\centering
\subfloat[\label{fig:4dsim-2_oct-a}]{%
\includegraphics[trim=1truemm 1truemm 2truemm 1truemm, width=0.49\columnwidth,clip=]{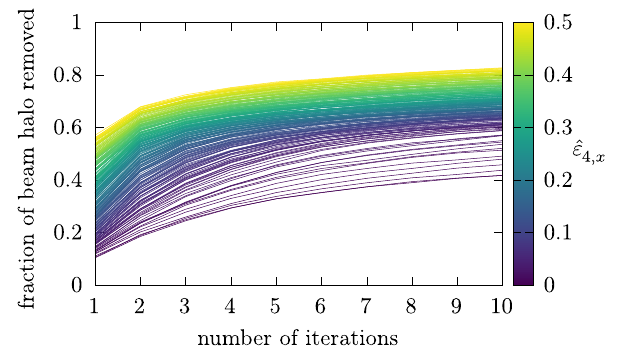}} 
%\hspace{10truemm} %new
\subfloat[\label{fig:4dsim-2_dec-a}]{%
\includegraphics[trim=1truemm 1truemm 2truemm 1truemm, width=0.49\columnwidth,clip=]{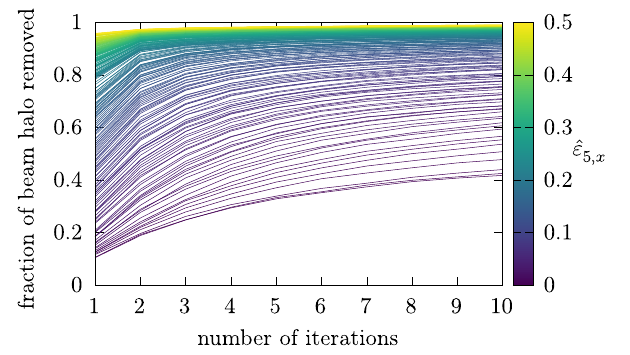}} \\ %new
\subfloat[\label{fig:4dsim-2_oct-b}]{% 
\includegraphics[trim=1truemm 1truemm 2truemm 1truemm, width=0.49\columnwidth,clip=]{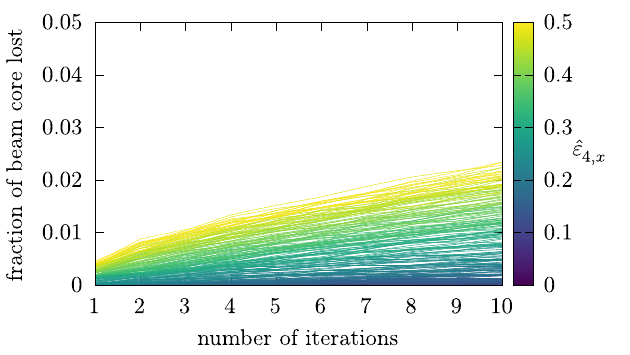}} 
%\hspace{10truemm} % new
\subfloat[\label{fig:4dsim-2_dec-b}]{%
\includegraphics[trim=1truemm 1truemm 2truemm 1truemm, width=0.49\columnwidth,clip=]{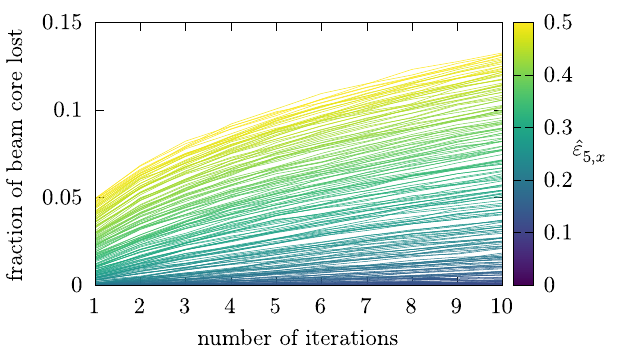}} \\ % new
\subfloat[\label{fig:4dsim-2_oct-c}]{%
\includegraphics[trim=1truemm 3truemm 5truemm 1truemm, width=0.49\columnwidth,clip=]{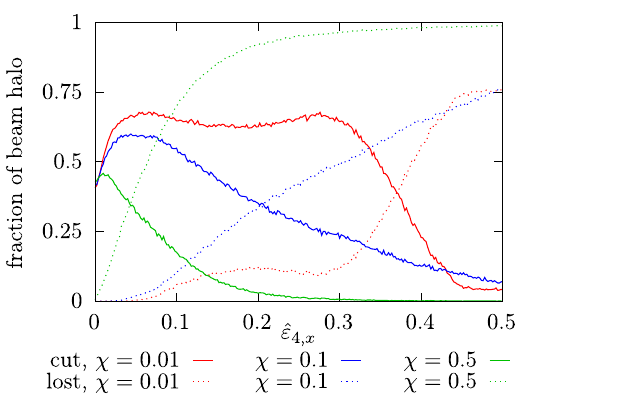}} 
%\hspace{10truemm} 
\subfloat[\label{fig:4dsim-2_dec-c}]{% %new
\includegraphics[trim=1truemm 3truemm 5truemm 1truemm, width=0.49\columnwidth,clip=]{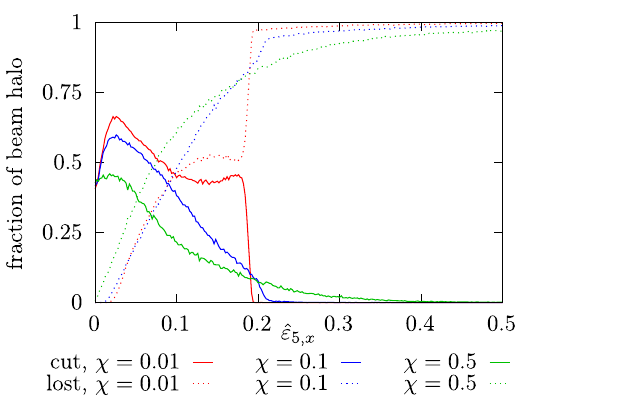}}  %new
\caption{Detailed studies of the beam halo cleaning procedure using the 4D model of the map in Eq.~\eqref{eq:map1}, and the configuration of Eq.~\eqref{eq:kicks1},  with $q=4$ (left column) or with $q=5$ (right column). We probe the fraction of beam halo removed (first row) and the fraction of beam core lost (second row) as a function of the number of iterations of the cleaning procedure for different values of $\hat\eps_q$ (encoded on the color scale). We also plot in the third row the fraction of beam halo removed in a controlled way (cut) or lost on the dynamic aperture (lost) as a function of the coupling strength $\chi$ (third row). We follow the protocol of Fig.~\ref{fig:process} (bottom); unless otherwise stated, we set $\hat k_3=1$, $\hat k_4=0$, $\hat\varepsilon_{q,y} = 5\hat\varepsilon_{q,x}$, $\chi=0.1$, $\omega_x=0.414$, $\omega_y=0.424$, $\omega_{x,\text{i}}=0.413$, $\omega_{x,\text{f}}=0.407$, $\omega_{x,\text{f}}=0.423$, $\omega_{y,\text{f}}=0.417$, $N=\num{1e5}$ and the beam-halo cleaning process is repeated $10$ times.} \label{fig:4dsim-2}
\end{figure*}

Figure~\ref{fig:4dsim-2} provides further insight into the characteristics of the halo-cleaning protocol. The plots in the left and right columns correspond to the octupole and decapole cases, respectively. The two upper rows display the fraction of the halo removed (top) and the fraction of the core removed (center) as functions of the number of repetitions of the trapping and transport stages. This analysis is performed with respect to the strength $\hat\eps_{q,x}$ of the AC multipoles. Overall, the decapole proves to be more effective than the octupole, although the octupole outperforms the decapole at low AC-multipole strengths. Additionally, the fraction of halo removed saturates more rapidly at higher multipole strengths. The middle plots reveal the behavior of core losses, which limits the permissible range of AC-multipole strength and the number of protocol repetitions.

Halo particles can be removed through either the trapping and transport mechanism, which moves them to the $r_2$ threshold, or by pushing them towards the system's dynamic aperture, reaching the boundary of stability. Both mechanisms are useful for our purposes, although the first offers a more controlled approach. These phenomena are analyzed using dedicated post-processing of numerical simulation data, with results depicted in the plots at the bottom of Fig.~\ref{fig:4dsim-2}. The proportion of halo particles that are cut (removed by trapping and transport) or lost (pushed to the stability boundary) is plotted against $\hat\eps_{q,x}$ for some values of $\chi$. Naturally, a stronger multipole results in a higher fraction of lost particles. For the decapole case, the transition from cut to lost particles is abrupt, whereas for the octupole case, the transition is smoother.

\begin{figure*}
\centering
\includegraphics[trim=1truemm 0truemm 3truemm 2truemm, width=0.49\columnwidth,clip=]{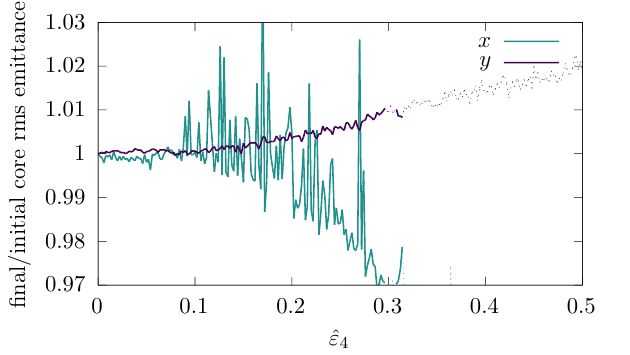} \includegraphics[trim=1truemm 0truemm 3truemm 2truemm, width=0.49\columnwidth,clip=]{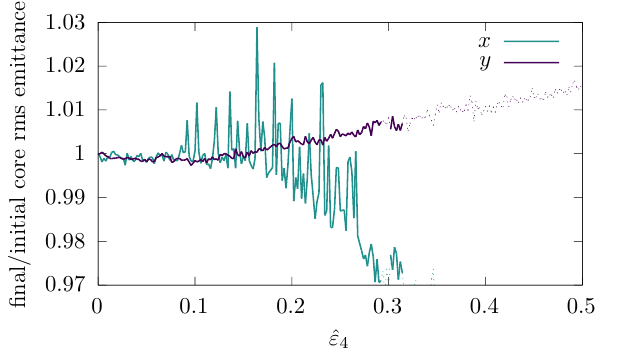}\\ %new
\includegraphics[trim=1truemm 0truemm 3truemm 2truemm, width=0.49\columnwidth,clip=]{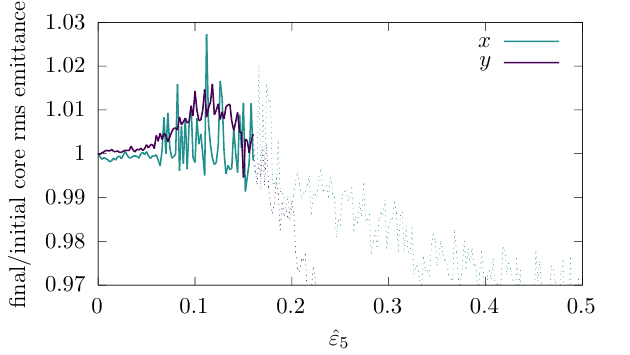} \includegraphics[trim=1truemm 0truemm 3truemm 2truemm, width=0.49\columnwidth,clip=]{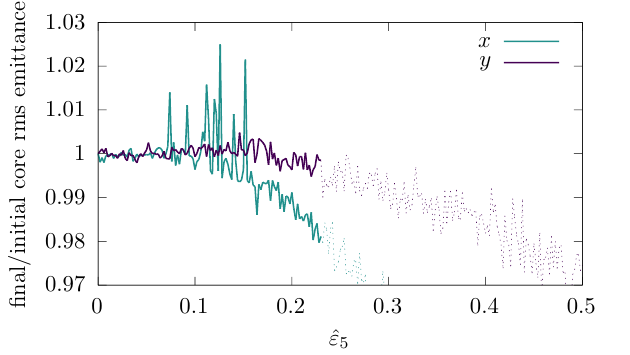}
\caption{Ratio between final and initial rms emittance in the $x$ and $y$ planes for the octupole (top) and decapole (bottom) model as a function of $\hat\eps_q$ in the 4D map model of Eq.~\eqref{eq:map1} and the configuration of Eq.~\eqref{eq:kicks1} (left column) and the map with the configuration of Eq.~\eqref{eq:kicks2} (right column). The initial emittance is defined as the beam rms emittance after $\num{1e6}$ turns performed with $\hat\eps_q=0$. This allows computing the impact of the filamentation due to nonlinearities used to generate the amplitude detuning. Note that the filamentation effects generate an increase in the rms emittance of $0.56\%$ in the $x$ plane and a $0.08\%$ reduction in the $y$ plane compared to the rms emittance of the ensemble of initial conditions for the configuration of Eq.~\eqref{eq:kicks1}, and a $0.52\%$ ($x$) and $0.31\%$ ($y$) increase for the configuration of Eq.~\eqref{eq:kicks2}. We used $\chi=0.1$, $\hat\eps_{q,y} = 5\hat\eps_q$, $\hat\eps_{q,x}=\hat\eps_q$ $\omega_x=0.414$, $\omega_y=0.424$, $\omega_{x,\text{i}}=0.413$, $\omega_{x,\text{f}}=0.407$, $\omega_{y,\text{i}}=0.423$, $\omega_{y,\text{f}}=0.417$, $N=\num{1e5}$.}
\label{fig:4dsim_emitt}
\end{figure*}

Finally, Fig.~\ref{fig:4dsim_emitt} (left column) illustrates the variation in the $x$ and $y$ rms emittances in the octupole (top) and decapole (bottom) scenarios, as a function of the AC multipole strength $\hat\eps_q$. The rms emittance values at the conclusion of the beam-halo removal process are normalized to their initial values. Similarly to the 2D case, the emittance growth caused by filamentation due to amplitude detuning terms in the 4D map has been assessed and used for normalizing the emittance values. Notably, the horizontal rms emittance utilized for normalizing the numerical results shows a growth by a factor of $1.0056$ relative to the initial core emittance, while the vertical plane exhibits a slight reduction ($0.9992$). The trends are comparable for both scenarios, with the horizontal emittance typically decreasing and the vertical emittance generally increasing. Halo cleaning with an octupolar element shows a marginally broader range of strength that maintains nearly constant emittance values. It is important to note that fluctuations in the horizontal emittance are caused by a few outliers in the particle distribution.

\subsection{4D map model: symmetric amplitude detuning}
The amplitude detuning serves as a crucial parameter in the model evaluating the proposed halo-cleaning method. Consequently, along with the analysis of asymmetric amplitude detuning between the horizontal and vertical planes, an additional analysis has been conducted for the scenario where the amplitude detuning is identical in both transverse planes. This is modeled with normal and skew nonlinear elements to produce the amplitude detuning terms, as indicated by Eq.~\eqref{eq:kicks2}. A primary result of this choice is that the strength of the AC multipoles has been equalized, which means $\hat{\eps}_{q,x}=\hat{\eps}_{q,y}=\hat{\eps}_q$.

\begin{figure*}
\centering
\subfloat[\label{fig:4dsim-3-oct-a}]{%
\includegraphics[trim=1truemm 1truemm 3truemm 2.5truemm, width=0.49\columnwidth,clip=]{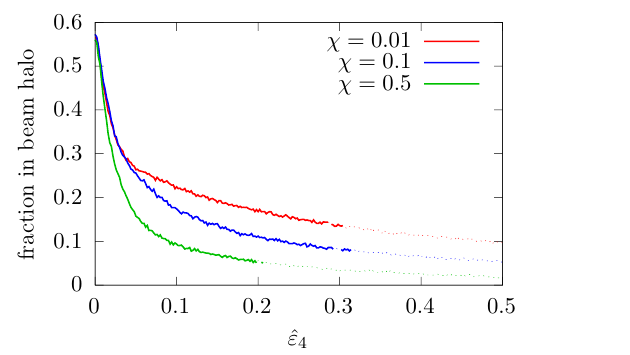}} %\hspace{20truemm} % nuova figura inserita
\subfloat[\label{fig:4dsim-3-dec-a}]{%
\includegraphics[trim=1truemm 1truemm 3truemm 2.5truemm, width=0.49\columnwidth,clip=]{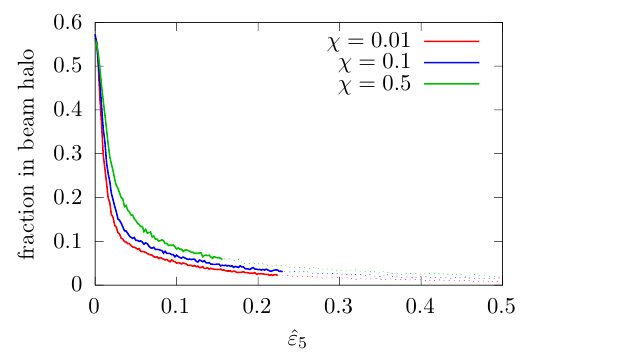}} \\ % nuova figura inserita
\subfloat[\label{fig:4dsim-3-oct-b}]{%
\includegraphics[trim=0truemm 3truemm 3truemm 13truemm, width=0.49\columnwidth,clip=]{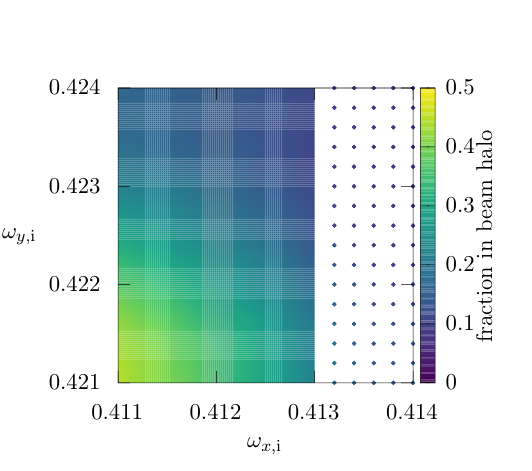}} %\hspace{20truemm}
\subfloat[\label{fig:4dsim-3-dec-b}]{%
\includegraphics[trim=0truemm 3truemm 3truemm 13truemm, width=0.49\columnwidth,clip=]{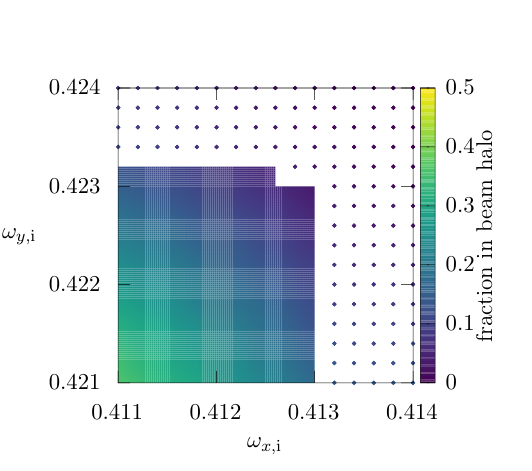}} \\
\subfloat[\label{fig:4dsim-3-oct-c}]{%
\includegraphics[trim=1truemm 3truemm 3truemm 13truemm, width=0.49\columnwidth,clip=]{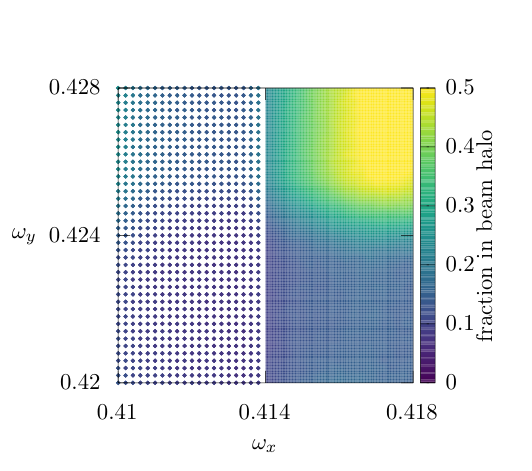}} %\hspace{20truemm}
\subfloat[\label{fig:4dsim-3-dec-c}]{%
\includegraphics[trim=1truemm 3truemm 3truemm 13truemm, width=0.49\columnwidth,clip=]{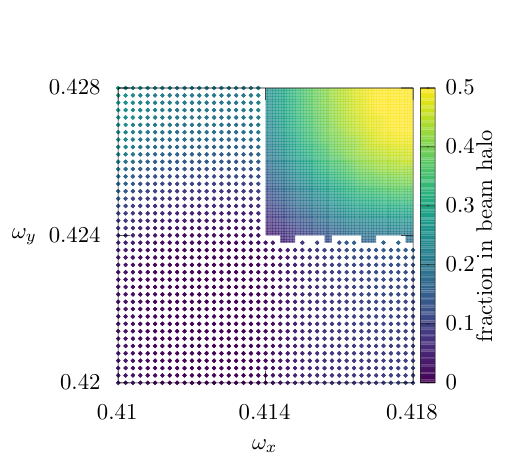}}
\caption{Survival fraction of particles of the beam halo after the cleaning procedure using the 4D model of the map in Eq.~\eqref{eq:map1}, and the configuration of Eq.~\eqref{eq:kicks2}, with $q=4$ (left column) or with $q=5$ (right column). Unless otherwise stated, we used $\hat k_3=1$, $\hat k_4=0$, $\hat\varepsilon_{q} =0.1$, $\hat\varepsilon_{q,y} = 5\hat\varepsilon_{q,x}$, $\chi=0.1$, $\omega_x=0.414$, $\omega_y=0.424$, $\omega_{x,\text{i}}=0.413$, $\omega_{x,\text{f}}=0.407$, $\omega_{x,\text{f}}=0.423$, $\omega_{y,\text{f}}=0.417$, $N=\num{1e5}$ and the beam-halo cleaning process is repeated $10$ times. In the 2D histograms, the filled regions represent conditions that preserve more than $99\%$ of the beam-core intensity.}
\label{fig:4dsim-3}
\end{figure*}

The results of the numerical simulation are shown in Fig.~\ref{fig:4dsim-3}, where the case with an octupole AC multipole is shown in the left column, while that with a decapole AC multiples is shown in the right column. 

The first feature that appears in comparison to Fig.~\ref{fig:4dsim} is the improvement in the performance of the halo-cleaning method, which is mainly due to the symmetric detuning. This has a clear impact as it reduces the asymmetry in the nonlinear coupling between the two transverse planes. The greatest improvement is obtained for the octupole case. Once again, the decapole outperforms the octupole also for the symmetric case depicted in Fig.~\ref{fig:4dsim-3}. These results indicate that for a model in which the nonlinear effects in the two transverse planes are balanced, the performance of the proposed method to clean the beam halo can be extremely successful, easily achieving a cleaning level in excess of 90\%.

The other significant finding refers to the effect of the cleaning method on the core beam emittance, illustrated in Fig.~\ref{fig:4dsim_emitt} (right column). Here, too, the symmetric model allows the cleaning method to perform optimally. Notably, the emittance growth is less pronounced than that with the asymmetric model, and this holds true across a wider range of AC multipole strengths.

To conclude, we emphasize that for the scenarios analyzed, specifically the asymmetric and symmetric amplitude detuning, the strength ratio of the AC multipoles employed to clear the beam halo has been set to specific optimal values: a factor of five for the asymmetric scenario and unity for the symmetric one. However, in a real-world circular accelerator lattice, an intermediate situation might occur, where the amplitude detuning in the two transverse planes is similar but not identical. In such instances, the strength ratio of the AC multipoles should be adjusted to regain optimal method performance, which has proven to be effective.
\section{Conclusions} \label{sec:conc}
This paper proposes and thoroughly examines a new method for beam-halo cleaning based on adiabatic trapping into nonlinear resonances, supported by theoretical analysis and comprehensive numerical simulations.

The technique employs AC multipoles to trap and transport particles in the beam halo in phase space. This approach signifies a novel advancement compared to the AC dipole cooling method applied to annular beam distributions. The key feature of our approach is the use of higher-order AC multipoles rather than dipoles, thereby preventing potential emittance growth in the beam core caused by AC dipoles.

An analytical study of a 2D model is presented to elucidate the theoretical principles of the technique, and numerical simulations demonstrate its promising potential. In addition, two more realistic 4D models have been extensively analyzed through numerical simulations. The primary distinction between these models lies in the amplitude detuning, which varies between being equal or different in the two transverse planes. The effectiveness of the method is significantly influenced by this condition, although it consistently achieves impressive halo cleaning results (around 60\%-70\%). When the amplitude detuning is equal in both the horizontal and vertical planes, the halo removal efficiency reaches 90\%.

A comprehensive assessment of the procedure was carried out, evaluating the sensitivity to various parameters such as the duration of the trapping and transport stages, the number of repetitions of these stages, the system's tunes, and the frequencies and strengths of the AC multipoles. In each case, ranges of parameter values were identified that ensured good performance. 

This paper's theoretical analyses, along with the positive outcomes from the numerical simulations, indicate that the suggested method is a promising strategy for addressing the major problems caused by the beam halo in modern and upcoming high-energy circular hadron colliders. Nonlinear AC multipoles can potentially trap and direct particles within the beam halo to high amplitudes, allowing interception by the collimator system. This strategy can be implemented whenever the beam halo becomes overly populated due to various processes in the accelerator, surpassing a set threshold.

Further studies are certainly necessary to understand the interplay of the halo formation mechanisms and the adiabatic resonance transport. For example, if the halo formation is driven by a diffusion process due to a large weak chaotic region in the phase space, one has to study the effectiveness of an adiabatic trapping mechanism in the presence of weak chaos in the phase space.

Potential avenues for further research include the examination of the dynamics generated by AC multipoles in realistic accelerator lattice configurations and the evaluation of the practicality of these magnetic elements, along with an initial assessment of their potential hardware performance.

\clearpage
\appendix 

\section{Detailed derivation of the 2D Hamiltonian model} \label{sec:app}
To determine the power of $J$ that enters in the Hamiltonian for a $1:1$ resonance driven by an AC magnetic multipole, we start applying the perturbation theory to the Hamiltonian of Eq.~\eqref{eq:ham_xp} written using the coordinates $x=\sqrt{2I}\cos\theta$, $p=\sqrt{2I}\sin\theta$, namely
\begin{equation}
\ham_0(\theta,I)=\omega_x I + \frac{2^{3/2} \beta^{3/2}_x \, k_3}{3}I^{3/2}\cos^3\theta
\end{equation}
to find new action-angle coordinates $(\phi,J)$ that recasts the Hamiltonian in the Normal Form
\begin{equation}
\hat{\ham}_0(J) = \omega_x J + \sum_l \frac{\Omega_l}{l}J^l\,.
\end{equation}

To this aim, we use the method of Lie transformations using a generating function $F(\phi,J)$ of the form 
\begin{equation}
F(\phi,J)= \beta^3_x \, k_3^2\sum_m f_m(\phi) J^{m/2}
\end{equation}
and we determine perturbatively $f_m(\phi)$ and $\Omega_l$ solving the equation 
\begin{equation}
\exp(D_{F(\phi,J)}) \ham_0(\theta,I) = \hat{\ham}_0 (J)
\label{eq:PVZ}
\end{equation}
in a pertubative way in the powers of $J$. $D_F \ham_0$ is the Lie derivative defined by the Poisson bracket $\{\ham_0, F\}$, and its $n$th power is defined recursively as $D^n_F \ham_0 = D_F \left (D^
{n-1}_F \ham_0 \right ) = \{\{\cdots\{\ham_0,F\},\cdots,F\},F\}$.

At each order, Eq.~\eqref{eq:PVZ} results in a differential equation for $f_m(\phi)$, the first terms being:
\begin{equation}
\begin{split}
f_3(\phi) & =-\frac{2 \sqrt{2} \sin \phi}{9 \, \beta^{3/2}_x \, k_{3} \omega_x} \left( \cos^2 \phi + 2 \right) \\
f_4(\phi) & = \frac{\cos \phi \sin \phi}{12 \, \omega_x^{2}} \left ( 2 \, \sin^2 \phi + 3 \,  \right ) + \\ 
          & + \phi \frac{{\left(5 \, \beta^3_x \, k_{3}^{2} + 6 \, \omega_x \, \Omega_{2}\right)}}{12 \, \beta^3_x \, k_{3}^{2} \, \omega_x^{2}}\, .
\end{split}
\end{equation}

The quantity $\Omega_2$ is then found by eliminating the secular term from $f_4$:
\begin{equation}
\Omega_2 = -\frac{5\, \beta^3_x \, k_3^2}{6\,\omega_x}\, .
\end{equation}

We can then proceed to the subsequent terms, which read
\begin{equation}
\begin{split}
f_5(\phi) & = \frac{\sqrt{2} \beta^{3/2}_x \, k_3 \sin \phi }{3240 \, \omega_x^{3}} {\left(132 \, \sin^4 \phi + 55 \,  \sin^2 \phi - 1785 \,  \right)} \\
f_6(\phi) & =
 \frac{2 \, \beta^3_x \, k_3^2 \sin \phi \cos \phi}{3888 \,  \omega_x^{4}} \, \times \\
 & \times \left(62 \, \cos^4 \phi - 1298 \, \cos^2 \phi + 1395 \right) + \\
 & + \phi \frac{9 \, {\left(235 \, \beta^6_x \, k_{3}^{4} + 144 \, \omega_x^{3} \, \Omega_{3}\right)}}{3888 \, \beta^3_x \, k_{3}^{2} \omega_x^{4}} 
\end{split}
\end{equation}
and once again the secular term can be set to zero if
\begin{equation}
\Omega_3= -\frac{235 \, \beta^6_x \, k_3^4}{144 \, \omega_x^3}\, .
\end{equation}

Finally, we compute
\begin{equation}
\begin{split}
f_7(\phi) & = \frac{\sqrt{2} \beta^9_x \, k_{3}^{3} \sin\phi}{93312 \, \omega_x^{5}} \left ( 9216 \sin^6 \phi - 9552 \sin^4 \phi  + \right . \\
& - \left . 8275 \sin^2 \phi  - 98295 \right ) \, ,
\end{split}
\end{equation}
which prevents us from going further, increasing the order of the Poisson brackets, solving the differential equations for $f_m(\phi)$ and determining $\Omega_l$ by canceling the secular terms in $f_m(\phi)$.

The next step is to compute the transformation of the AC-exciter term in the Hamiltonian, namely $\ham_q(\theta, I, t)$, which is given by 
\begin{equation} 
\ham_q(\theta,I,t) = \eps_q \frac{(2I)^{q/2}}{q}\cos^q\theta \cos\omega t
\end{equation}
to the $(\phi,J)$ coordinates using the generating function $F(\phi, J)$. When averaging over time, the only non-vanishing contributions are the terms of the Fourier expansion of $x^n =  (2I)^{q/2}\cos^q\theta$ corresponding to the components $\exp(\pm i\phi)$.

If $q$ is odd, it is easy to see that the lowest-order, non-vanishing contribution to the Fourier series is given by the term
\[ 
(2J)^{q/2}\cos^q \phi 
\]
since
\begin{equation}
\begin{split}
    \frac{1}{2\pi}\int_0^{2\pi}\dd\phi\, e^{i\phi} \cos^q\phi & = \frac{1}{2^q}\binom{q}{\frac{q-1}{2}} \\
    & = \frac{q!}{2^q \qty(\frac{q-1}{2})! \qty(\frac{q+1}{2})!}
\end{split}
\end{equation}
and therefore the Hamiltonian reduces to
\begin{equation} 
\hat{\ham}_q (\phi,J,t) = \frac{\eps_q}{2^{q/2}} \frac{(q-1)!}{ \qty(\frac{q-1}{2})! \qty(\frac{q+1}{2})!} J^{q/2}\cos\phi\cos\omega t\, .
\end{equation}

If $q$ is even, the situation is more complicated: the term immediately given by the linear part of the transformation has a null projection on $\cos\phi$, therefore a further inspection to the higher orders of the transformation is needed, namely up to $J^{\hat m/2}$, with $\hat m \ge q/2$.

Depending on $q$, we retrieve different values of $\hat m$, which result in different exponents in the AC part of the Hamiltonian. For a sextupole, the first non-zero term in the Fourier expansion appears at $q=3$, and we have
\begin{equation}
c_{3,3}=\frac{\sqrt{2}}{4}
\end{equation}
while for an octupole ($q=4$) and a decapole ($q=5$) the resonant component appears at $J^{5/2}$, and we have
\begin{equation}
c_{4,5}=\frac{7\sqrt{2} \, \beta^{3/2}_x \, k_3}{12 \, \omega_x}\,,\qquad c_{5,5}=\frac{\sqrt{2}}{4} \, ,
\label{eq:Fourier}
\end{equation}
respectively. We can continue with a dodecapole and a $14$-pole to find in both cases $q=7$ (note that in that case the Hamiltonian will also include a term $\propto \Omega_3 J^3$) and
\begin{equation}
c_{6,7}=\frac{35\sqrt{2} \, \beta^{3/2}_x \, k_3}{44\omega_x}\,,\qquad c_{7,7}=\frac{5\sqrt{2}}{16} \, ,
\end{equation}
respectively. In general, we can conjecture that $\hat m = q+1$ if $q$ is even and $\hat m = q$ if $q$ is odd: a proof should look in detail into the trigonometric terms in the generating function of the transformation. 
\section{Fixed points of the Hamiltonian with an AC multipole} \label{sec:app1}
In Section~\ref{sec:gen}, the analysis of the Hamiltonian of Eq.~\eqref{eq:hamxy} has been carried out, also considering the existence and properties of the fixed points on the coordinate axis. Here, we will show that there are no other fixed points to be considered. To show this, we consider the general equations whose solutions are given by the fixed points, namely the system of equations
\begin{equation}
\left \{ \begin{aligned}
    \pdv*{\ham}{X} & = 0 \\
    \pdv*{\ham}{Y} & = 0
\end{aligned} \right .
\label{eq:syst0}
\end{equation}
or
\begin{equation}
\left \{ \begin{aligned}
     2 \lambda X + 4 X \left ( X^2 + Y^2 \right ) + 4 \mu_q X^2 \left ( X^2 + Y^2 \right ) + \\
     + \mu_q \left ( X^2 + Y^2 \right )^2 & = 0 \\
     2 \lambda Y + 4 Y \left ( X^2 + Y^2 \right ) + 4 \mu_q X Y \left ( X^2 + Y^2 \right ) & = 0 \, ,
\end{aligned} \right .
\label{eq:syst1}
\end{equation}
which is a system with $16$ complex solutions, if counted with their multiplicity, and four solutions are given by the fixed points on the $X$-axis. Hence, there are $12$ more solutions. Furthermore, we note that if $(\bar{X}, \bar{Y})$ is a real solution of Eqs.~\eqref{eq:syst0}, then by symmetry, also $(\bar{X}, -\bar{Y})$ is a solution, which indicates that real solutions appear in pairs. If $(\tilde{X}, \tilde{Y})$ is a complex solution of Eqs.~\eqref{eq:syst0}, then $(\tilde{X}^\ast, \tilde{Y}^\ast)$, the complex conjugate of $(\tilde{X}, \tilde{Y})$, is a solution of Eqs.~\eqref{eq:syst0} because it is a system of polynomial equations with real coefficients. However, the invariance of the equations by the symmetry $Y \to -Y$ still holds, which means that complex solutions appear in quadruples. In the end, the following situations are possible: 3 quadruples of complex solutions; 2 quadruples of complex solutions and two pairs of real solutions; 1 quadruple of complex solutions and 4 pairs of real solutions.

The system~\eqref{eq:syst0} can be recast in a more convenient form by transforming to a polar coordinate system $(\delta, r)$ defined as $X=r \cos\delta$, $Y=r \sin\delta$, namely
\begin{equation}
\left \{ \begin{aligned}
     2 \lambda r \cos\delta + 4 r^3 \cos\delta + 4 \mu_q r^4 \cos^2\delta + \mu_q r^4 & = 0 \\
       \lambda r \sin\delta + 2 r^3 \sin\delta + 2 \mu_q r^4 \cos\delta \sin\delta & = 0 \, ,
\end{aligned} \right .
\label{eq:syst2}
\end{equation}
from which one obtains the following equations
\begin{equation}
\left \{ \begin{aligned}
    \mu_q r^6 & - 4 r^5 + 4 r^4 + 4 \vert \lambda \vert r^3 - 4 \vert \lambda \vert r^2 - \lambda^2 r + \lambda^2 = 0 \\
     \cos \delta & = - \frac{\lambda +2 r^2}{2 \mu_q r^3} \, ,
\end{aligned} \right .
\label{eq:syst3}
\end{equation}
where we have used the assumptions that $\lambda < 0$ and $r>0$. 

The first equation can be studied by considering the two functions
\begin{equation}
\begin{split}
    f(r) & = \mu_q r^6 \\
    g(r) & = \left (r-1 \right ) \left ( \sqrt{2} r - \sqrt{\vert \lambda \vert} \right )^2 \left ( \sqrt{2} r + \sqrt{\vert \lambda \vert} \right )^2 \, ,
\end{split}
\label{eq:syst-funct}
\end{equation}
so that the solutions of the first equation of the system~\eqref{eq:syst3} are the solutions of $f(r) = g(r), \, r > 0$. 

It is straightforward to observe the following properties of $g(r)$, namely
\begin{itemize}
\item $g(r)=0$ for $r_1=\sqrt{\vert \lambda \vert/2}$ and $r_3=1$.
\item $g(r)>0$ for $r> r_3$.
\item $g(r)$ has a maximum for $r_1=\sqrt{\vert \lambda \vert/2}$ and a mi\-ni\-mum for $r_2=\frac{4+\sqrt{16+10 \lambda^2}}{10}$. For the case under consideration, i.e.. $\lambda \lesssim 0$, we have $r_2 < r_3$.
\end{itemize}

These observations allow to state that:
\begin{itemize}
\item If $\mu_q < 0$ then the first equation of the system~\eqref{eq:syst3} has a single solution $\bar{r}< r_1$.
\item If $\mu_q > 0$, then the first equation of the system~\eqref{eq:syst3} has no real solution because $f(r)>0, \, \forall r>0$ and $f(r_1)=\mu_q> 0 =g(r_1)$ and $f(r)/g(r) \to +\infty$ for $r\to +\infty$.
\end{itemize}
\bibliographystyle{unsrt}
\bibliography{mybibliography, extrabib}
\end{document}